%% file: 0-main.tex
\documentclass[sigconf]{acmart}

\newcommand \tool{\texttt{UIGuard}}
\newcommand{\red}{\textcolor{black}}

\newcommand{\jieshan}{\textcolor{teal}}

\AtBeginDocument{%
  }

\copyrightyear{2023} 
\acmYear{2023} 
\setcopyright{acmlicensed}\acmConference[UIST '23]{The 36th Annual ACM Symposium on User Interface Software and Technology}{October 29-November 1, 2023}{San Francisco, CA, USA}
\acmBooktitle{The 36th Annual ACM Symposium on User Interface Software and Technology (UIST '23), October 29-November 1, 2023, San Francisco, CA, USA}
\acmPrice{15.00}
\acmDOI{10.1145/3586183.3606783}
\acmISBN{979-8-4007-0132-0/23/10}

\usepackage{array,graphicx}
\usepackage{float}
\usepackage{multirow}
\usepackage{makecell}
\usepackage{soul}
\usepackage{subfig}
\usepackage{graphicx}

\usepackage{tikz}

\newcommand*{\adTriInfoIcon}{\includegraphics[scale=0.2]{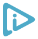}}%
\newcommand*{\adCloseIcon}{\includegraphics[scale=0.2]{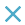}}%

\begin{document}

\title{Unveiling the Tricks: Automated Detection of Dark Patterns in Mobile Applications}

\author{Jieshan Chen}
\orcid{0000-0002-2700-7478}
\email{Jieshan.Chen@data61.csiro.au}
\affiliation{%
  \institution{CSIRO's Data61}
  \country{Australia}
}

\author{Jiamou Sun}
\orcid{0000-0002-5212-7068}
\email{Frank.Sun@data61.csiro.au}
\affiliation{%
  \institution{CSIRO's Data61}
  \country{Australia}
}

\author{Sidong Feng}
\orcid{0000-0001-7740-0377}
\email{Sidong.Feng@monash.edu}
\affiliation{%
  \institution{Monash University}
  \country{Australia}
}

\author{Zhenchang Xing}
\email{Zhenchang.Xing@data61.csiro.au}
\authornote{Also with Australian National University.}
\orcid{0000-0001-7663-1421}
\affiliation{%
  \institution{CSIRO's Data61}
  \country{Australia}
}

\author{Qinghua Lu}
\email{Qinghua.Lu@data61.csiro.au}
\orcid{0000-0002-7783-5183}
\affiliation{%
  \institution{CSIRO's Data61}
  \country{Australia}
}

\author{Xiwei Xu}
\email{Xiwei.Xu@data61.csiro.au}
\orcid{0000-0002-2273-1862}
\affiliation{%
  \institution{CSIRO's Data61}
  \country{Australia}
}

\author{Chunyang Chen}
\email{Chunyang.Chen@monash.edu}
\affiliation{%
  \institution{Monash University}
  \country{Australia}
}

\renewcommand{\shortauthors}{Jieshan Chen et al.}

\begin{abstract}
Mobile apps bring us many conveniences, such as online shopping and communication, but some use malicious designs called dark patterns to trick users into doing things that are not in their best interest. Many works have been done to summarize the taxonomy of these patterns and some have tried to mitigate the problems through various techniques. However, these techniques are either time-consuming, not generalisable or limited to specific patterns. 
To address these issues, we propose \tool{}, a knowledge-driven system that utilizes computer vision and natural language pattern matching to automatically detect a wide range of dark patterns in mobile UIs. Our system relieves the need for manually creating rules for each new UI/app and covers more types with superior performance. 
In detail, we integrated existing taxonomies into a consistent one, conducted a characteristic analysis and distilled knowledge from real-world examples and the taxonomy. Our \tool{} consists of two components, Property Extraction and Knowledge-Driven Dark Pattern Checker. 
We collected the first dark pattern dataset, which contains 4,999 benign UIs and 1,353 malicious UIs of 1,660 instances spanning 1,023 mobile apps. Our system achieves a superior performance in detecting dark patterns \red{(micro averages: 0.82 in precision, 0.77 in recall, 0.79 in F1 score)}. A user study involving 58 participants further shows that \tool{} significantly increases users' knowledge of dark patterns. 
\end{abstract}






\maketitle

\input{1-intro}

\input{2-relatedWork}

\input{3-charateristics}

\input{4-approach}
\input{5-dataset.tex}

\input{7-experiments.tex}

\input{8-user_study.tex}
\input{9-discussion}
\input{10-threats}
\input{11-conclusion}

% ...

\bibliographystyle{ACM-Reference-Format}
\bibliography{reference}


\appendix
\input{12-appendix}
% ...

\end{document}

%% file: 1-intro.tex
\section{Introduction}
\label{sec:intro}

\begin{figure*}
    \centering 
    \includegraphics[width=0.98\textwidth]{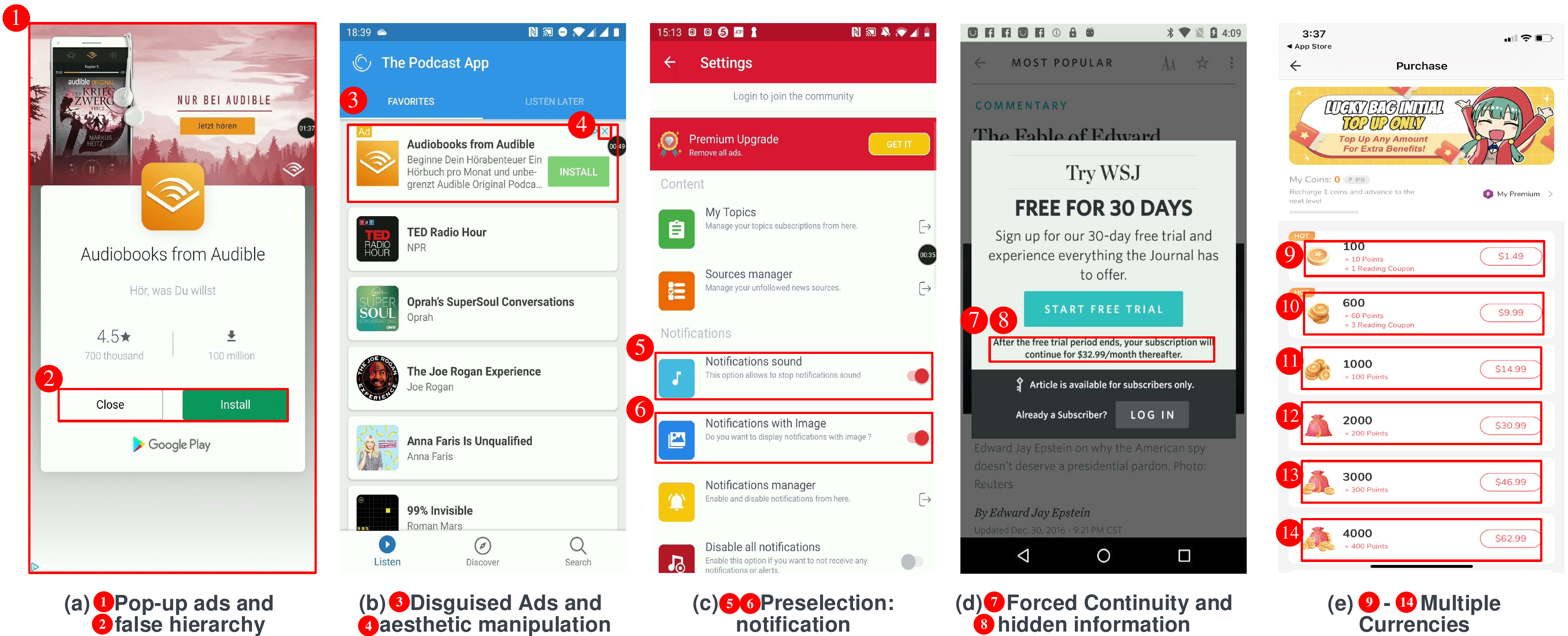}%
    \Description{From left to right. (a) A screenshot displaying an ad page UI with two distinct buttons, ``Close'' and ``Install'', centrally located.
    (b) A UI screenshot showing podcast channels listed alongside an ad, identical to regular content, and featuring a small, possibly challenging-to-click close button on the top right.
    (c) A UI showcasing a settings page with the notification setting enabled.
    (d) A UI displaying a pop-up window encouraging users to sign up for a 30-day free trial, with small text at the bottom indicating automatic continuation of subscription for \$32.99/month post-trial.
    (e) A payment page UI where users can purchase virtual currency with real currency.}
    \caption{Examples of dark patterns in mobile apps. \red{(a) A screenshot displaying an ad page UI with two distinct buttons, ``Close'' and ``Install'', centrally located.
    (b) A UI screenshot showing podcast channels listed alongside an ad, identical to regular content, and featuring a small, possibly challenging-to-click close button on the top right.
    (c) A UI showcasing a settings page with the notification setting enabled.
    (d) A UI displaying a pop-up window encouraging users to sign up for a 30-day free trial, with small text at the bottom indicating automatic continuation of subscription for \$32.99/month post-trial.
    (e) A payment page UI where users can purchase virtual currency with real currency. }
    }
    \label{fig:examples}
\end{figure*}

Mobile apps have been indispensable parts in our daily life, bringing conveniences and essential functionalities in various aspects, such as online shopping, remote work, and communication with families and friends. 
However, the user interfaces (UIs) of these apps, which serve as a bridge between end-users and the app, can sometimes contain malicious design elements\footnote{https://www.deceptive.design/hall-of-shame/all}, that may cause harms to end-users.
One specific example of this is the use of ``dark patterns'',  which are maliciously crafted user interfaces designed to trick users into doing something they do not want to do~\cite{brignull2010dark}. These dark patterns can take many forms and cause different types of damage.
For example, preselection tricks (see Figure~\ref{fig:examples}(c)) automatically select favorable options for the app, pressuring users to subscribe to unwanted newsletters. Roach motel patterns make it easy for users to sign up for a service but difficult or impossible for them to unsubscribe, causing frustration and inconvenience~\cite{brignull2010dark, gray2018dark}.
Recent research found that ~11\% of ~11K shopping websites contain dark patterns~\cite{mathur2019dark}, and 95\% of 240 popular mobile apps include at least seven different types of malicious designs in their UIs on average~\cite{di2020ui}.

To raise the awareness of dark patterns and categorise them, many researchers conducted different systematic empirical studies on understanding and evaluating the existence of dark patterns in various platforms~\cite{mathur2019dark,gray2018dark, di2020ui} and contexts~\cite{nguyen2022freely, gray2021dark, aagaard2022game, gak2022distressing, habib2022identifying}.
In 2010, Harry Brignull created a website called dark pattern~\cite{brignull2010dark}.
He not only built a website that catalogued and defined 12 types of dark patterns on internet and provided illustrative examples, but also continuously collects and exposes new dark pattern instances to ``name and shame'' companies through crowd-sourcing techniques on Twitter~\cite{twitterDark}.
Following Brignull, many studies have expanded our understanding of dark patterns and their implementation in user interfaces by identifying additional strategies and types, analyzing their potential harm, and conducting a comparative study. They shed light on the prevalence and impact of dark patterns on user behavior and decision-making in various contexts~\cite{gray2018dark, di2020ui, mathur2019dark, gunawan2021comparative}.
However, these taxonomies are heterogeneous and lack consistency, which creates obstacles for classifying specific dark patterns. 

In parallel, to mitigate and overcome the existence of dark patterns, some researchers proposed various solutions, including crowd-sourcing techniques~\cite{twitterDark, redditDark}, developer/user patches~\cite{kollnig2021want}, or using some naive text-based classification techniques~\cite{soe2022automated,insite}.
However, these techniques are either time-consuming, hard to generalise or limited to certain patterns.
Crowd-sourcing techniques aim to provide a platform for end-users to report instances of dark patterns in their daily life and raise awareness~\cite{twitterDark, redditDark}, but they are time-consuming and may not be generalised to new UIs. 
People who do not read these posts are unable to learn this knowledge.
Developer patches involve modifying code files to remove dark patterns and repackaging apps~\cite{kollnig2021want}, but this method requires programming expertise and may not scale well. Moreover, patches can accidentally affect app functionality, and repackaging apps can raise privacy concerns~\cite{datta2022greasevision}.
User patches allows users to provide screenshots of areas to be modified, but this approach requires significant user input and is not generalizable.~\cite{datta2022greasevision}.
Text-based classification techniques can automatically detect and highlight/remove partial dark patterns, but they require a large labelled dataset to train the model and may produce many false positives and false negatives as they only rely on texts~\cite{soe2022automated}.  
Overall, these solutions have limitations in terms of scalability, generalizability, and effectiveness, and there is a need for further research to develop more effective methods for detecting and addressing dark patterns.

In this work, we take a step to unify the existing taxonomies and propose a knowledge-driven system called \tool{} that automatically detects dark patterns in mobile UIs.
Our system leverages computer vision and natural language matching techniques to effectively and efficiently detect dark patterns, covers a wider range of types with superior performance and relieves the need for creating rules or patches for each new UI/app. 
To achieve it, we first analysed and merged existing taxonomies of dark patterns in mobile platforms~\cite{di2020ui, gray2018dark, gunawan2021comparative} into a consistent one to form a solid knowledge base of our detection system. 
We then conducted a characteristic analysis on the unified taxonomy to inform the design of our automated dark pattern detection system at the screen level and element level, and identified identified six essential properties, including element coordinates and types, text contents, icon semantics, element status, element colors and relationships, for detecting dark patterns.

Our \tool{} leverages computer vision and natural language pattern matching techniques.
It is purely vision-based, taking only a UI screenshot as input and requiring no other metadata or information.
This enables the generalisability of our proposed techniques to other platforms.
\tool{} consists of two steps, namely UI element property extraction and knowledge driven dark pattern checker.
UI element property extraction involves seven modules to gradually extract vision and textual properties from a UI screenshot. 
Knowledge driven dark pattern checker incorporated the heuristics distilled from our knowledge base. 

To evaluate the efficiency and usefulness of our proposed systems, we use Rico dataset~\cite{deka2017rico} and its semantic labeling~\cite{liu2018learning}, as our base datasets, to train the deep learning modules.
To evaluate the whole system, we established a set of standards to annotate the types and locations of dark patterns. We selected the testset of Rico dataset as our annotation target, which includes 6,352 UIs of 1,023 apps from 27 app categories. 
The annotation process, carried out by two authors in two steps, resulted in 4,999 benign UIs and 1,353 malicious UIs. These malicious UIs contained a total of 1,660 instances of 14 types of dark patterns. 
The results show that \tool{} can accurately and efficiently detect dark patterns in mobile UIs, with an accuracy of 0.93 for classifying whether a UI contains dark patterns or not.
The overall performance for detecting and recognizing the types of dark patterns \red{achieves 0.83/0.82 in precision, 0.82/0.77 in recall and 0.82/0.79 in F1 score for the macro/micro average respectively.}
Our ablation experiments in Section~\ref{sec:ablation}, which evaluate the impacts of each module, further demonstrates the validity and necessity of each module.

Finally, we carried out a user study with 58 participants of diverse background, which validates that our tool could significantly help users to gain more knowledge on dark patterns, with the recall rate increasing from 18.5\% to 57.8\%.
The participants highly appreciate \tool{}, and see its great potential in helping users, companies and regulators.
We also explore the limitations and future directions of our work. 
\red{We also note that it is crucial to consider the potential for misuse of the tool rigorously before incorporating them into regulatory practices.}

Our contributions can be summarised as:
\begin{itemize}
    \item We conducted the first systematic analysis of existing dark pattern taxonomies on mobile apps, integrated existing taxonomies and carried out a characteristic analysis on them, which enlightens the design of our automated dark pattern detection tool.

    \item We developed the first dark pattern detection tool, \tool, which adopts computer vision and natural language pattern matching techniques to distill information from user interfaces and identify dark patterns in mobile UIs.
    
    \item We constructed the first large-scale dark pattern dataset\red{~\cite{uiguardDataset}}, which contains 4,999 benign UIs and 1,353 malicious UIs of 1,660 instances spanning 1,023 mobile apps. All source code and datasets can be found at our GitHub repository~\footnote{https://github.com/chenjshnn/UIST23-UIGuard}.

    \item We conducted a user study with 58 participants, which verifies the usefulness of our tool and the potential educational usage.
\end{itemize}

%% file: 2-relatedWork.tex
\section{Related Work}
Our work is related to three topics, i.e., taxonomies of dark patterns, dark pattern detection and UI understanding,

\subsection{Taxonomies of Dark Pattern }
There are many existing empirical studies on the definition of dark patterns in various platforms, including websites and mobile apps~\cite{mathur2019dark, gray2018dark, di2020ui}, and in different domains, such as game~\cite{aagaard2022game}, social media platforms~\cite{habib2022identifying}, shopping websites~\cite{mathur2019dark}, cookie banner~\cite{nguyen2022freely, gray2021dark} and advertisements~\cite{gak2022distressing}. 
The general workflow of existing empirical studies start at exploring and collecting user interfaces of desktop or mobile applications, and researchers then summarize the types of dark patterns they identified in the user interfaces.

In 2010, Harry Brignull~\cite{brignull2010dark} created the dark pattern website, defining dark patterns as ``tricks used in websites and apps that make you do things that you didn't mean to, like buying or signing up for something''. 
He also created a twitter account~\cite{twitterDark} for people to report and discuss the dark patterns they see in their daily usage.
Later, Gray et. al~\cite{gray2018dark} manually collected 112 artifacts from popular online platforms for a two-month period, and extends Brignull's concept by summarizing the concepts in terms of five strategies, i.e., \textbf{nagging, obstruction, sneaking, interface interference and forced action}, and adding more dark patterns subtypes and examples.
Based on these taxonomies, Mathur et al.\cite{mathur2019dark} conducted a large-scale empirical study on around 11K shopping websites.
They collected the data by simulating user shopping behaviours and clustering segments in collected websites.
They manually checked the clusters and summarized the dark patterns existing in top shopping websites, and recognised 1,818 instances of dark patterns from around 11\% shopping websites.
They analysed five characteristics (asymmetric, covert, deceptive, hides information and restrictive) of dark patterns and six cognitive bias (anchoring, bandwagon, default, framing effects, scarcity bias and sunk cost fallacy), and investigated how dark patterns leverage these cognitive bias to affect user behaviours.
Different from Mathur et. al's focus on shopping websites, Di Geronimo et. al~\cite{di2020ui} evaluated the dark pattern issues on 240 popular Android apps by recording 10-min usage videos for each app.
They reused the taxonomies from Gray et. al~\cite{gray2018dark} and identified some new dark pattern types (16 types and 31 cases).
The taxonomies become more finer-grained and more complete when Gunawan et. al~\cite{gunawan2021comparative} conducted a comparative study of dark patterns across different modalities (mobile and web) in terms of the phase of usage. 

In this paper, we focus on detecting dark patterns in mobile apps but we need a solid and consistent knowledge base of dark pattern as our foundation. Therefore, we carefully examine existing taxonomies, and integrated them into a single and unified one.


\subsection{Dark Pattern Detection}
There are different strategies to \textbf{detect} dark patterns in the wild, including manual exploration, semi-automated clustering based methods and some naive text-based classification methods.

The detection of dark patterns started with some manual exploration, either by the domain researchers~\cite{gray2018dark, di2020ui, gunawan2021comparative, brignull2010dark} or the ordinary users~\cite{twitterDark, redditDark}.
After these researchers or ordinary users detect the potential dark patterns in some applications, they will either share their experience with the related UI screenshot, established a website or published some papers, as a way to help and educate other end-users to avoid being tricked and raise the awareness.
However, while the detection is of high accuracy, this method suffers from several drawbacks.
First, it is very time-consuming and require people to have enough dark pattern knowledge. 
Second, as apps keep updating and new apps emerges over the time, such manual detection could not capture all these changes.

To mitigate the shortages of manual exploration, Mathur et al.~\cite{mathur2019dark} proposed semi-automatic techniques to simulate user behaviours and identify dark patterns in shopping websites.
They adopted the clustering technique to ease the detection process by grouping related UI patterns, and then manually evaluated the clusters instead of going through each artifact one by one.
As a result, they were able to analyse ~53k product pages from ~11k shopping websites.
While their approach eases the process of manual exploration, it is still time-consuming and hard to apply to new UIs.

Based on their findings and dataset, some~\cite{insite, soe2022automated} considered using classification techniques to detect these dark patterns in the wild.
For example, Tung et al.~\cite{insite} leveraged the text data from Mathur et al.~\cite{mathur2019dark} and train a Bernoulli Naive Bayes based classifier for websites, and Soe et al.~\cite{soe2022automated}
developed a gradient boosted tree classifier to identify the dark patterns in cookie banners.
However, these techniques are all text-based and limited to certain dark patterns and may easily introduce many false positives as we analysed in Section~\ref{sec:analysis} and Section~\ref{sec:ablation}.
They also require a labelled training dataset of dark pattern to train their models.

Our research focuses on developing a system that can automatically detect dark patterns in mobile applications by incorporating domain knowledge. Our approach offers several advantages over previous work in this area. Firstly, our methods are scalable and can be used to analyze dark patterns across large numbers of apps. Additionally, our system is designed to monitor new UIs as they are released, ensuring that it remains effective even as the landscape of mobile applications evolves. Compared to existing approaches, our tool offers superior coverage and performance, and our modular method allows easy integration of new rules and property identification modules for new types of dark patterns and platforms. By promoting awareness of these unethical design practices and encouraging responsible usage of mobile apps, we hope to prevent the manipulation of users through the implementation of dark patterns.

\subsection{UI Understanding}
Understanding the semantic of user interfaces can assist many downstream tasks, such as accessibility testing and enhancement~\cite{chen2020unblind, liu2020owl, chen2022towards, zhang2021screen, chen2022extracting}, automated UI testing~\cite{white2019improving}, UI design search~\cite{chen2020wireframe, li2021screen2vec}, conversational agents~\cite{li2020multi} and UI code generation~\cite{chen2018ui, chen2020object}. 
The detection of dark patterns also require the understanding of UI from different perspectives, such as UI element detection and icon understanding.

While previous research has explored the semantic understanding of UIs for various tasks, no existing work has focused specifically on detecting dark patterns. In our study, we conducted a comprehensive analysis of the characteristics of dark patterns, allowing us to extract richer information from UIs. In addition to the tasks mentioned in previous work, we also considered factors such as contrasting colors to measure visual saliency, analyzing text content, and understanding element status. Our approach is the first to enable automated dark pattern detection in mobile UIs, covering 14 different types of dark patterns and achieving superior performance.
Our system, \tool{}, is designed to provide clear explanations of its decision-making process, giving end-users a better understanding of how it operates and the accuracy of its decisions. Through our work, we hope to promote awareness of the prevalence of dark patterns in mobile apps and prevent their unethical usage. By providing a reliable and transparent tool for detecting these patterns, we aim to empower users to make informed decisions about the apps they choose to use. 



%% file: 3-charateristics.tex
\section{Background}
Dark patterns have been well-investigated and defined by many researchers, and many taxonomies from different angles are been proposed. 
However, these taxonomies lack consistency and are analyzed from different perspectives. 
In this work, we focus on detecting dark patterns in mobile applications; therefore, we first built a solid and consistent knowledge base to guide the design of our automated dark pattern detection system.
We based on three existing taxonomies related to mobile apps \cite{gray2018dark, gunawan2021comparative, di2020ui}, and integrated them in Table~\ref{tab:taxonomy}, which shows five strategy categories, types and subtypes of dark patterns, definitions and cases, as our basic taxonomy..
With their provided definitions and related examples, we can better understand and find the root and essence of these dark patterns. 
We first merged some cases in \cite{gunawan2021comparative, di2020ui} into one as they are highly similar.
For example, \cite{gunawan2021comparative} has three separate cases ``Optional add-on items are preselected'',  ``Consent notice includes email/SMS subscriptions with a preselected opt-in checkbox'' and ``checkbox options are preselected''.
We merged them into one single case `` notification/subscriptions or other options are preselected'' as the rationale of detecting these cases are the same.
We also removed some website related cases as we focus on mobile apps. For example, we removed ``the cookie consent is preselected''.
However, as we include the preselection type, the cookie consent one can actually be considered as a special case of it.
Note that our technique could be easily generalised to these websites related cases, but we first focus on mobile UIs to demonstrate the potential of our approach.

\input{tables/taxonomy.tex}
Specifically, Gray et al. summarized five categories of dark patterns as seen in Table~\ref{tab:taxonomy}.
\begin{itemize}
    \item \textbf{Nagging} is defined as a repeated app action that interrupts the users' current task and nags them to do something else. Examples are like popping up an irrelevant windows for advertisements or ratings. 

    \item \textbf{Obstruction} makes things harder to do that favours the app company's interest. There are three types, namely \textit{Roach Motel} (easy to get in, hard to get out), \textit{Price Comparison Prevention}(unable to select product names to make comparisons in other platforms) and \textit{Intermediate Currency} (disconnect users from real money by using virtual currency).

    \item \textbf{Sneaking} tries to hide, disguise or delay information that is relevant to users. It has four types. \textit{Forced Continuity} forces users to auto continue the service when their  purchase one expires. \textit{Hidden Costs} delays information like high tax rate or delivery fee at the late stage of checkout. \textit{Sneak into Basket} quietly adds additional items to users without their consent. \textit{Bait and Switch} misleads users to action by giving false expectations.
    
    \item \textbf{Interface Inference} privileges some options over others to confuse and hide the information from users. It contains four types. \textit{Hidden Information}, which diminishes the visibility of information relevant to users. \textit{Preselection} selected unfavorable options by default. \textit{Aesthetic Manipulation} leverages visual effects to distract/attract users' attention from/to specific options or actions. It includes four sub-types, \textit{Toying with emotion} (e.g., countdown offer to make users nervous), \textit{False Hierarchy} (specific options are more prominent than others), \textit{Disguised Ad} (Ads looks the same as other content) and \textit{Tricked Questions}(e.g., double negation questions).

    \item \textbf{Forced Action} forces users to perform some actions to get rewards, unlock features or achieve some tasks. \textit{Social Pyramid} asks users to leverage their contacts to get rewards, \textit{Privacy Zuckering} made users to share more information that are not necessary, and \textit{Gamification} grinds users by repeated tasks to get some rewards.

\end{itemize}


\section{Analysis}
\label{sec:analysis}
To better lay the foundation for the automated dark pattern detection tool for mobile apps, we conducted a characteristic analysis on these dark patterns at the screen level and element-level. 

\subsection{Screen-level perspective}
Based on the unified taxonomy, we classified these dark patterns into two categories: static dark patterns and dynamic dark patterns.

A static dark pattern means that the dark pattern is specific to a single user interface and does not depend on the context or history of the user's interactions. 
They can be identified by analyzing the current UI, without the need to consider the user's past actions.
This may involve extracting information about the elements on the UI, such as their coordinates, types, colors, text content, and icon semantics, as well as the relationships between these elements. 
For example, in Figure~\ref{fig:examples}(a), to detect the \texttt{False Hierarchy} dark pattern, we might group the ``Close'' and ``Install'' buttons together and compare their text and background colors to determine if the button that favors the app provider's benefit is more prominent. 
Another example of a static dark pattern is \texttt{Disguised Ads}.
As seen in Figure~\ref{fig:examples}(b), it requires us to first identify the advertisement block by extracting the text content of the badge and locating the ad info icon, and understand this block is visually similar to other contents.
This example also involve a \texttt{General Aesthetic Manipulation} dark pattern as the close button on the top-right of the ad block is very small, and the end-user may easily wrongly click the ad and trigger an unwanted AD page, wasting their time and bringing bad user experience.
Understanding the semantic and size of icon is vital for examining this dark pattern.
There are also many other dark pattern types fall into this category, such as \texttt{Intermediate Currency} and \texttt{Forced Continuity}.
We marked static dark patterns as \checkmark  in the ``Is Context-Independent?'' column of Table~\ref{tab:taxonomy}.

\begin{figure*}
    \centering
    \includegraphics[width=1.0\textwidth]{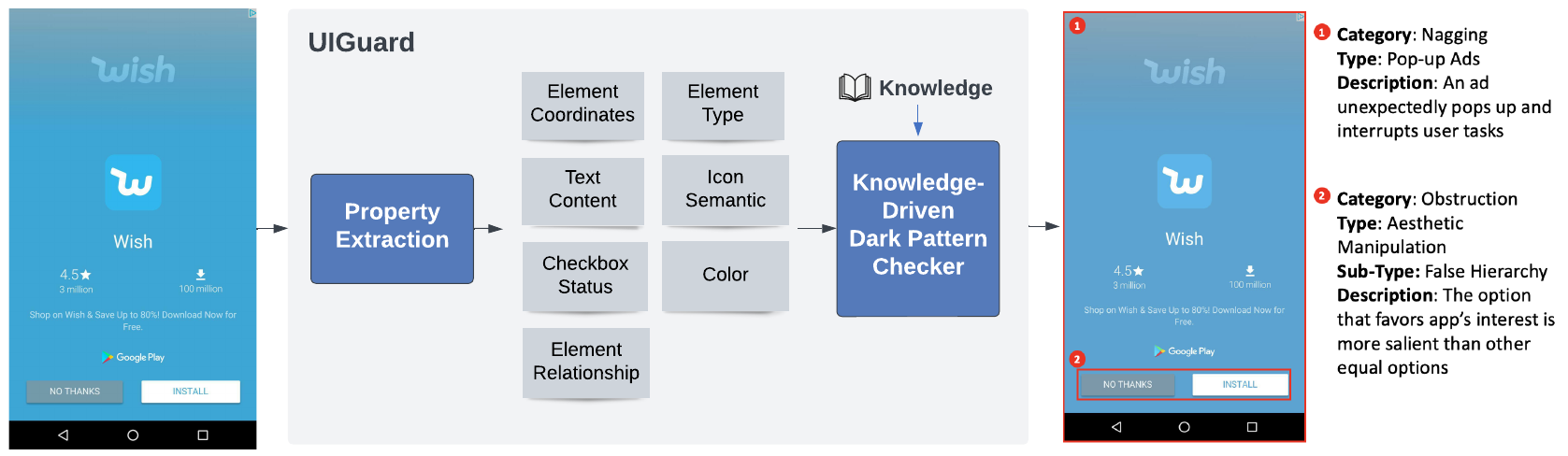}
    \caption{Overview of our approach. \red{Given a UI screenshot, our Property Extraction module extracts element coordinates and types, text content, icon semantic, element status, background and foreground colors and element relationship. Then, powered by knowledge, our Knowledge-Driven Dark Pattern Checker identifies potential dark patterns,  delivering detection results and highlighting these patterns with a red box.}
    }
    \Description{This figure shows the workflow of our proposed system. Given a UI screenshot, our Property Extraction module extracts element coordinates and types, text content, icon semantic, element status, background and foreground colors and element relationship. Then, powered by knowledge, our knowledge-driven Dark Pattern Checker identifies potential dark patterns,  delivering detection results and highlighting these patterns with a red box.}
    \label{fig:approach}
\end{figure*}

In comparison, dynamic dark patterns are context-dependent and require additional information beyond the current user interface in order to be identified. This may include previous UIs, the elements that triggered the current page or animation effect, and the user's intent.
For example, in order to verify the existence of a \texttt{Bait and Switch} dark pattern, we need to understand the user's intent and determine whether the current UI meets their expectations. 
Similarly, \texttt{Hidden Costs} dark patterns can be detected by comparing previous UIs to see if the app revealed these costs beforehand. 
On the other hand, \texttt{Roach Motel} dark patterns require an analysis of all UIs within the app to determine whether there are options for logging out, unsubscribing, or deleting the account. 
In our analysis, for dynamic dark patterns, we denoted $\times$ in the ``Is Context-Independent?'' column of Table~\ref{tab:taxonomy}.


Some dynamic dark patterns can be identified based on the current UI, even though they are context-dependent.
One example is \texttt{Nagging}, which involves repeated interruptions of the user's task through pop-up windows that ask the user to rate or upgrade the app, or watch an advertisement (as shown in Figure~\ref{fig:examples}(a)).
These UIs typically appear without being triggered by the user, and their occurrence suggests a high possibility of a dark pattern.
In this case, we can warn users about the potential for dark patterns. 
Another example is \texttt{Preselection}, which involves determining whether an option is selected by the user or enabled by default(see Figure~\ref{fig:examples}(c)).
It may not be possible to determine this information even with contextual UIs, so we also consider it an ``in-between'' type that warrants a warning to users.
Users may choose to turn off the detection of this dark pattern type if they do not need the warnings.
For this type of dark pattern, it is necessary to understand the element type and status, as well as the text content.
In our analysis, we marked these ``in-between'' dark patterns with ``Hybrid'' in Table~\ref{tab:taxonomy}.

There is one type of dark pattern that cannot be identified solely from the UI: \texttt{Price Comparison Prevention}. This type of dark pattern requires interaction with the elements on the UI in order to determine if texts can be copied. To detect this pattern, we need to obtain instrumental information about the copyability of certain elements. In our analysis, we marked this dark patterns with a ``/'' in Table~\ref{tab:taxonomy}.

In our paper, we presented the first systematic approach to detecting dark patterns within a single UI. As such, we focus specifically on static and ``in-between'' dark patterns, which can be identified based on the current UI.

\subsection{Element-level perspective}

Dark patterns can be related to a single element or a combination of elements. 
To identify single element dark patterns, we can evaluate the properties of each element individually, such as its coordinates, contained text, colors, icon semantics, and element type. 
For example, \texttt{Forced Continuity} dark patterns can be detected by examining the text content of an element, as in Figure~\ref{fig:examples}(d). 
It contains a text element saying ``7 days free, then \$84.00/year'', which automatically opt-in users after the free trail ends. 
The size and color of the text may also be relevant for identifying \texttt{Hidden Information} dark patterns.

Multiple element dark patterns can be further divided into two categories. One type requires analyzing the attributes of several elements to confirm the presence of a dark pattern, such as the co-existence of virtual and real currency in Figure~\ref{fig:examples}(e), which indicates the presence of \texttt{Intermediate Currency} issues. The other type requires a higher level of information, such as the relationship between UI elements. For instance, Figure~\ref{fig:examples}(a) shows an example of a \texttt{False Hierarchy} dark pattern, which requires grouping related buttons (such as the accept and cancel buttons) and comparing their visual saliency to determine if one option is privileged. This process involves understanding the relative position and characteristics of the elements.


\subsection{Characteristic Identification}

Our analysis of dark patterns has identified six key characteristics that are necessary for detecting these patterns within user interfaces. 
These include element meta information such as \textit{coordinates and element types}, which can help identify the shape and relationships of elements; 
the \textit{text content} of elements, which is useful for determining the semantics of certain patterns; 
the \textit{status} of certain types of elements, such as checkboxes; 
\textit{icon semantics}, which can sometimes be used to identify the overall meaning of a UI page or block; background and text \textit{colors}, which are relevant for visual patterns like False Hierarchy; and higher-level information such as the grouping of similar elements and the \textit{relationships} between UI elements, which is important for multiple element patterns. 
By considering all of these factors, we can develop a comprehensive and effective strategy for detecting dark patterns.


%% file: tables/taxonomy.tex
\begin{table*}[]
    \caption{Integrated taxonomy of Dark Patterns. From left to right, it shows the category, types, sub-types, and their descriptions. Following these, we show the characteristics analysis results. Hybrid in Column ``Is Context-Independent?'' means that the current case is context related but can easily be spotted by a single UI page.}
    \vspace{-10pt}
    \centering
    \resizebox{0.90\textwidth}{!}{
    \begin{tabular}{p{0.15\linewidth}  p{0.2\linewidth} p{0.2\linewidth} p{0.25\linewidth} p{0.25\linewidth}|c c|c|c c c c c | c}
     \textbf{Category}  & \textbf{Type} & \textbf{Sub-Type} & \textbf{Description} & \textbf{Cases} & \rotatebox[origin=c]{90}{\textbf{Is Context-Independent?}} & \rotatebox[origin=c]{90}{\textbf{Related to Multiple Elements?}} & \rotatebox[origin=c]{90}{\textbf{Our Tool}} 
     & \rotatebox[origin=c]{90}{\textbf{Coordinates \& Type }} 
     & \rotatebox[origin=c]{90}{\textbf{Text Content}} 
     & \rotatebox[origin=c]{90}{\textbf{Element Status}}  
     & \rotatebox[origin=c]{90}{\textbf{Icon Semantic}} 
     & \rotatebox[origin=c]{90}{\textbf{Color}}
     & \rotatebox[origin=c]{90}{\textbf{\# of Instance in Rico Testset}}
     \\ 
    \hline
    \multirow[t]{4}{\linewidth}{\textbf{NAGGING (NG)}} 
    & & & \multirow[t]{4}{\linewidth}{A pop-up window unexpectedly and repeatedly appears and interrupts user tasks.} 
          & Pop-up ads        & hybrid & $\times$ & $\checkmark$ & $\times$ & $\checkmark$ & $\times$ & $\checkmark$ & $\times$ & 178 \\ %
    & & & & Pop-up to rate    & hybrid &$\times$&$\checkmark$& $\times$ & $\checkmark$ & $\times$ & $\checkmark$ & $\times$ & 1 \\ %
    & & & & Pop-up to upgrade & hybrid &$\times$&$\checkmark$& $\times$ & $\checkmark$ & $\times$ & $\times$     & $\times$ & 9 \\ %
    \hline
    \textbf{OBSTRUCTION (OB)}
    & Roach Motel & & Easy to opt-in, not possible or hard to opt out & logout$/$unsubscribe $/$delete account &$\checkmark$ & /  &$\times$&   &   &   &   &   &  \\
    & Price Comparison Prevention & & Hard to make direct comparison with other markets & Cannot copy product names while shopping & \slash & $/$  & $\times$ &   &   &   &   &   & \\
    & Intermediate \newline Currency & & Disconnect users with the real money by asking them purchase virtual currencies & Buy virtual coins for gaming $/$ reading & $\times$ &$\checkmark$ &$\checkmark$& $\times$ & $\checkmark$ & $\times$ & $\times$ & $\times$ & 0 \\
    \hline
    \textbf{SNEAKING (SN)}
    & Forced \newline Continuity & & Users continue to be charged after the service expired. & Free trial and auto opt-in & $\times$ &$\times$&$\checkmark$& $\times$ & $\checkmark$ & $\times$ & $\times$ & $\times$ & 1 \\
    & Hidden Costs & &  Late disclosure of certain costs & Tax $/$ delivery fee is shown at the final payment page. &$\checkmark$ & $/$ &$\times$&  &  &  &  &  & \\
    & Sneak into Basket & & Add items not chosen by the users &  additional items like gift cards auto added to the cart. &$\checkmark$  & $/$ &$\times$&  &  &  &  & & \\
    & Bait and Switch & &  You click a feature but gets undesired results & click a normal button, but get an ads &$\checkmark$ & $/$ &$\times$&  &  &  &  &  & \\
    \hline
    \multirow[t]{7}{\linewidth}{\textbf{INTERFACE \newline INTERFERENCE (II)}} 
    & \textbf{Hidden \newline Information} &  & Options or actions relevant to users are not made immediately or readily accessible
    & Use greyed small texts to show relevant info (e.g., terms of service) &$\times$ & $\checkmark$ & $\times$ & & & & & & \\
    \cline{2-14}
    & \textbf{Preselection} & & Options are preselected & notifications$/$subscriptions or other options are preselected &hybrid & $\checkmark$ & $\checkmark$ & $\checkmark$ & $\checkmark$ & $\checkmark$ & $\times$ & $\times$ & 146 \\
    & & & No checkbox & If ToS$/$PP present, no consent checkbox is provided & $\times$ & $\checkmark$ & $\checkmark$ & $\checkmark$ & $\checkmark$ & $\times$ & $\times$ & $\times$  & 86 \\
    \cline{2-14}
    & \textbf{Aesthetic \newline Manipulation (AM)}
    & Toying with emotion  & use language, color, style to evoke an emotion to nag users doing something & Countdown offer $/$ rewards  & $\times$ & $\checkmark$ & $\checkmark$ & $\times$ & $\checkmark$ & $\times$ & $\checkmark$ & $\checkmark$ & 0 \\ 
    & & & & Confirm shaming &$\times$  & $\times$  & $\times$  & & & & & & \\ 
    &   & False Hierarchy  & One option is more salient than other equal options & Accept Button looks more salient than Reject button & $\times$ & $\checkmark$ & $\checkmark$ & $\checkmark$ & $\checkmark$ & $\times$ & $\checkmark$ & $\checkmark$  & 273\\ 
    &   & Disguised Ad  & Ads pretends to be normal content & A sponsored content$/$ad looks like a normal content or icons$/$buttons are ads, but it is not clear  &$\times$ & $\checkmark$ & $\checkmark$ & $\checkmark$ & $\checkmark$ & $\times$ & $\checkmark$ & $\times$ & 46 \\ 
    & & & & Ad with Interactive game & hybrid & $\checkmark$ & $\times$  & & & & & & \\ 
    &   & Tricked Questions  & Use confusing wordings to ask questions & double negation  &$\times$ & $\checkmark$ & $\times$  & & & & & & \\ 
    & & General Types 
    &  & Small close buttons  &$\times$ & $\times$  & $\checkmark$ & $\checkmark$ & $\times$ & $\times$ & $\checkmark$ & $\times$ & 684 \\
    & & & & Moving ads buttons  & $\checkmark$ & $\times$ & $\times$ & & & & & & \\
    \hline
    \textbf{FORCED \newline ACTION (FA)}
    & Social Pyramid  &  & Users are asked to share something with their friends to get rewards or unlock features. & Invite friend to get voucher  &$\times$ &$\times$&$\checkmark$& $\times$ & $\checkmark$ & $\times$ & $\times$ & $\times$ & 16 \\ 
    & Privacy \newline Zuckering &  & Users are forced to share more information &  send usage data by default, agree terms of service by default &hybrid  & $/$  &$\checkmark$& $\checkmark$ & $\checkmark$ & $\checkmark$ & $\times$ & $\checkmark$ & 117 \\ 
    & Gamification  &  & Ask users to do a same thing to get something & Daily rewards &$\times$ & $/$ &$\checkmark$& $\times$ & $\checkmark$ & $\times$ & $\times$ & $\times$  & 0 \\ 
    \cline{2-14}
    & General types  &  & & Countdown on ads & hybrid &$\checkmark$&$\checkmark$& $\times$ & $\checkmark$ & $\times$ & $\checkmark$ & $\times$ & 4 \\ 
    & & & & Watch Ads to unlock features or get rewards & $\times$ &$\times$&$\checkmark$& $\times$ & $\checkmark$ & $\times$ & $\times$ & $\times$ & 2 \\
    & & & & Pay to avoid ads & $\times$ &$\times$&$\checkmark$& $\times$ & $\checkmark$ & $\times$ & $\times$ & $\times$ & 97 \\
    \hline
    \end{tabular}}
    \label{tab:taxonomy}
\end{table*}

%% file: 4-approach.tex
\section{Approach}

Figure~\ref{fig:approach} shows the flowchart of our system, \tool{}, which takes a UI screenshot as input and reports the types and locations of recognized dark pattern instance in the screen.
Generally, \tool{} consists of two parts, namely Property Extraction and Knowledge-Driven Dark Pattern Checker.
The Property Extraction module is responsible to intelligently extract essential properties from the UI screenshot, and the Knowledge-Driven Dark Pattern Checker powered by the knowledge we distilled from the integrated taxonomy, can conduct analysis on these information and examine the types and location of dark pattern instances in the current UI.
As an example in Figure~\ref{fig:approach}, the user inputs a UI screenshot, and \tool{} identifies two dark pattern instances within that UI, highlights them with red boxes and provides detailed explanations.

\subsection{UI Element Property Extraction}
 
\subsubsection{UI Element Detection}
\label{sec:ui_localization}
The first step of our system is to recognize UI elements in a given UI screenshot.
In our work, we use Faster-RCNN~\cite{ren2015faster} to localize and recognize UI elements in UIs, and then merges the detection in this step with the results from Text Content Extraction module to form more accurate detections.


\subsubsection{Text Content Extraction}
To extract text content in UI, we adopt a mature OCR model, PaddleOCR~\cite{paddle}, which supports over 80 language, and achieves high accuracy.
We use their pretrained model for English language, and obtain the bounding boxes of text lines and text contents in the UI screenshot.
Instead of inputting the detected UI element from the previous step one by one, we directly feed the whole UI screenshot to the OCR engine to save computing time and to retrieve miss-detected text elements.
After we obtain the text lines (coodinates and text contents), we merges these text detections with the detections from FasterRCNN.

\begin{table*}
    \centering
    \caption{Example patterns we distilled from our knowledge base to serve as the knowledge in identifying dark patterns. The full list can be seen in \cite{uiguardDataset}.}
    \resizebox{0.95\textwidth}{!}{
    \begin{tabular}{p{0.25\linewidth}|p{0.7\linewidth}|p{0.3\linewidth}}
    \hline
    \textbf{Type}                &  \textbf{Text Patterns (partial rules)}  & \textbf{Need any other information?} \\
    \hline
    FA - General - PRO (pay to avoid ads) & ...remove/without/disable/block/no .. ad/ ads/ advertisement/ advertisements/ advertising/adverts ... & No \\
    FA - Watch to unlock features  & ...watch... ad/ads/advertisements/ advertisements/ advertising/ adverts... & No \\
    NG - Pop up to Rate     &  if you enjoy/like ... apps, ...rate...  &  Star icons \\
    NG - Pop-UP AD        & Ad Text                           & AD-related Icons, Coordinates \\
    II - Preselection        &   .. consent/agree/give consent/ accept ... terms/ privacy policy/ policies/agreement]...  & Checkbox is checked \\
    II - Aesthetic Manipulation - False Hierarchy & One of the options should contain texts like  No/no thanks/ close/next time/later/ not now/skip/cancel to ensure the malicious intent
   & Color \& Grouping \\
    \hline
    \end{tabular}
    }
    \label{tab:text_patterns}
\end{table*}

\textit{Merging with FasterRCNN detections:} 
UI elements normally do not overlap with each other except one case when an ImageView element is served as a background of other elements.
Therefore, we first sort the UI elements by their size in ascending order.
For each UI element, we calculate the interaction over union (IoU - $interaction area/unioin area$) values with each detected text line. If the IoU value over a predefined threshold, we consider it as a match, and add the corresponding text content to the UI element. 
As one UI element may contain several text lines, we allow one UI element to match with several text lines. However, one text line can only be matched with one UI element. After we iterate all UI element, the unmatched text lines will be considered as TextView elements that is missed by our UI Element Detection module.

In addition, we notice that the bounding boxes of text elements from UI Element Detection are not very precise while OCR engines can extract highly precise bounding boxes of the text lines. Leveraging this feature, we further refine the detected bounding boxes by replace the bounding boxes of UI text elements with the merged bounding box of the matched text lines.

\subsubsection{Icon Semantic Understanding}
\label{sec:icon_classication}
Understanding the semantics of icons are important to recognise the existence of dark patterns. For example, star icons will appear in the \texttt{Nagging} patterns, and the triangle Ad information icons will always appear in ad-related dark patterns.
Hence, for UI elements detected as ImageView or ImageButton elements, we trained a ResNet-18 model~\cite{he2016deep}
to classify the icons into finer semantic categories.

In addition, we notice that some advertisement-specific icons, such as \adTriInfoIcon and \adCloseIcon, are normally not included in the view hierarchy, which means that our localisation model is ``trained to miss them''.
However, these icons are necessary to detect ad-related dark patterns like \texttt{Disguised Ads}.
Actually, most advertisements will have the exact same ad-specific icons.
The reason is that many advertisement network companies\footnote{https://digitaladvertisingalliance.org/participating}, such as Google Inc., Facebook, Microsoft and AT\&T, join the adChoices\footnote{https://youradchoices.com/} self-regulatory program, which aims to establish and enforce responsible practices for online behavioural advertising and give consumers enhanced transparency and control. 
By joining this program, their advertisement is required to provide the adchoice icon, i.e., a triangle info icon \adTriInfoIcon, to indicate that the ads are recommended based on their online behaviours.
Therefore, we adopt the template matching technique~\cite{brunelli2009template} to find these missing ad-specific icons.


\subsubsection{Element Status recognition.}
Finer details on the status of UI elements are necessary as explored in Section~\ref{sec:analysis}.
Therefore, to recognise the status of UI elements, we trained another light-weighted ResNet-18~\cite{he2016deep} model.

\subsubsection{Color Extraction}

Color can impact the visual saliency to manipulate user perception.
To extract the background and text color of UI elements, we use color histogram to first obtain the most frequently color as the background color.
We then obtain the foreground color by computing the euclidean distance between the rest color with the background color, and find the one with farthest distance. The rationale inside is that we assume the designers will make the background and foreground colors having highest contrast ratio so that end-users could clearly separate the foreground content from the background.

\subsubsection{UI Element Grouping}
Apart from the properties of UI elements, their relationships are also important, especially for \texttt{False Hierarchy} tricks, which need to compare the visual saliency among related UI elements.
Normally, the app providers or designers will manipulate the color and size of buttons or text elements to highlight or dampen some options.
Therefore, we use density-based spatial clustering of applications with noise (DBSCAN) algorithm~\cite{ester1996density}, which is a density-based clustering algorithm, to perform elements grouping, and only consider text elements and buttons.
The advantages of this clustering algorithm is that we do not need to pre-define the number of clustering.

The basic idea is that based on the element features, it can group UI elements in a high-density region, while marking those UI elements in low-density region as outliers.
To define whether an area is of high-density or of low-density, we need to define a distance function and two thresholds, i.e., a distance threshold $\alpha$ and a density threshold $\beta$.
We consider four features, including element types, size (width and height), coordinates and their text content, based on which we compute the distance between any two elements, to group relevant UI elements. After relevant elements are grouped together, we examine their color to see if they have big visual differences.


\subsection{Knowledge-Driven Dark Pattern Checker}

To effectively detect dark patterns in mobile UIs, we utilized our integrated taxonomy of dark patterns and drew upon existing literature for examples and datasets. 
We specifically analyzed the partially released videos from Di Geronimo et al.~\cite{di2020ui}, UXP2 gallery\footnote{https://darkpatterns.uxp2.com/patterns/} from Gray et al.~\cite{gray2018dark} and the Deceptive Design collection~\cite{brignull2010dark} from Harry Brignull.
From this, we distilled key patterns that allow for automated detection of dark patterns in UIs.
We show partial patterns in Table~\ref{tab:text_patterns}, and a complete one can be seen in \cite{uiguardDataset}.
For instance, the \texttt{Pay to Avoid Ads}
pattern can be identified by searching for phrases like ``remove ads'' or ``disable advertising'', while the \texttt{FA - Watch to Unlock Features} pattern can be detected by searching for phrases like ``watch ad''. In addition to text content, the semantic meanings of icons can aid in the identification of many AD-related dark patterns, as the presence of certain icons or AD badges often indicates the presence of advertising. Therefore, combining the existence of ad indicators and the location of that block, we can recognise \texttt{NG - Pop up AD} tricks.

Some patterns also require an understanding of the status of UI elements. The \texttt{Preselection} pattern, in which the option favored by the app provider is selected by default, can be identified by searching for notification, privacy policy, and usage data-related text, and then checking if the corresponding checkbox is preselected. The \texttt{False Hierarchy} pattern can be detected by comparing the visual differences between a group of relevant elements, and evaluating the content of buttons.

Upon obtaining UI elements from previous steps, our Knowledge-Driven Dark Pattern Checker uses the knowledge to scan the UI elements in a given UI and identify any instances of these patterns. As seen in Figure~\ref{fig:approach}, \tool{} was able to accurately detect two dark patterns in the example UI: nagging ads and false hierarchy. This demonstrates the power and effectiveness of our approach. 



%% file: 5-dataset.tex
\section{Datasets and Implementation}
\label{sec:datasets}

In this section, we present the datasets used to train and evaluate our proposed system, \tool{}. First, we discuss the deep learning datasets used to train and test our deep learning modules. Then, we introduce the dark pattern dataset that we annotated and used to evaluate the overall performance of \tool{} in detecting dark patterns.


\subsection{Deep learning datasets and model training}
We used the Rico dataset~\cite{deka2017rico} as the basis for our experimental dataset. 
\red{While it is collected in 2016, it remains the most widely-used and high-quality dataset for Android UI designs~\cite{deka2021early}. It consists of 66,261 GUIs from 9,384 free Android applications and includes view hierarchy information for each UI, such as layout information and element attributes (e.g., element type, content, coordinates). While it lacks semantic descriptions, Liu et al.~\cite{liu2018learning} added explicit names to UI elements based on predefined heuristics.} 
They identified 25 UI component categories, such as Page Indicators, CARDs, Checkboxes, and Icons, through an interactive open coding process. These two datasets allow us to train our models for element localization, icon recognition, and status recognition with minor efforts. 
\red{Despite the absence of new datasets, they mostly built upon and enhanced the Rico dataset from other aspects, such as captioning UIs and removing noises~\cite{liu2018learning, li2022learning}.}


\subsubsection{Element Localisation Dataset.}
We utilized the Rico dataset to train our element localization model (i.e., FasterRCNN) as it provides bounding boxes and types for contained GUI elements. We selected 15 types of Android GUI elements (e.g., Button, ImageView, TextView) as our target objects\footnote{We provide the list of UI element and examples in the supplementary materials}. We filtered out any UIs that do not contain these elements, resulting in a dataset of 50,524 UIs. We splitted the dataset into training, validation, and testing sets (80:10:10) and ensured that UIs from one app are only included in a single split to prevent potential data leakage.


We initialised the model parameters using the pretrained model of COCO dataset~\cite{lin2014microsoft} and finetuned all parameters using the processed Rico dataset. 
The model was trained on a Nvidia 1080 GPU for 160 epochs with a batch size of 8, an initial training rate of 0.001, and weights are updated by an Adam optimizer\cite{kingma2014adam}. To remove duplicate detection, we applied non-max suppression to choose the most confident detection from the overlapping detection. We set the interaction over union (IoU) threshold to 0.5, and found the best confidence threshold using the validation dataset, which is 0.65.

\subsubsection{Icon Recognition Dataset.}
To train our icon recognition model, we used the annotations from Liu et. al~\cite{liu2018learning}.
In total, they identified 135 icon types and labelled 73,449 icons.
We removed duplicate icons by comparing the color histogram of icons, and considered icon types that appear at least 200 times in our training dataset.
Note that icon types that appear less than 200 times are considered as an additional Other type.
After that, we noticed some icons are noisy data or misclassified.
Therefore, we manually examined all data and removed these wrong data.
As a result, we had 81 target icon types, such as star and close icons, and one \textit{other} type\footnote{All icon types can be seen in supplementary materials.}.
Note that while for now, we only require the identification of several icons, we still train a general icon recognition model for future extension.
In addition, as the bounding boxes of detections may not be as accurate as the ground truth bounding boxes, we performed some data augmentation techniques to enhance the robustness.
Specifically, we randomly added some noises to the groundtruth bounding boxes to simulate the potential inaccurate detections.
All icons were splited into training, validation and testing dataset using the same method as the element localization dataset. This means that if an icon's UI is in the training dataset, the icon will also be placed in the training dataset.

We rescaled the icon to a input size of $224 \times 224$.
We initialised the ResNet-18 model parameter using the pretrained model on ImageNet dataset and finetuned it on a Nvidia 1080 GPU for 100 epochs with a batch size of 128 and an initial training rate of 0.001.
The model weights were updated by stochastic gradient descent optimizer.

\subsubsection{Status Recognition Dataset.} 
For our status recognition model, we extracted UI elements with a type of ``Checkbox'', ``Switch'' or ``Toggle Button'' from the semantic dataset. 
After we obtained the data, we then manually annotated the status of them as the Rico semantic dataset does not contain such information.
Same as the icon classification model, we also performed data augmentation by randomly adding some noises to the bounding boxes.
As a result, we obtained 1,845 checked elements and 1,383 unchecked elements.
As some non-checkbox elements may be wrongly detected as checkboxes, we trained a 3-class model to deal with the problem.
Therefore, we randomly extracted 3,093 other elements as a negative class for the model to mitigate the potential classification errors inherited from element localization step.
This dataset was splited following the same strategy as in previous datasets.

The model was trained by inputting a RGB image of input size $224 \times 224$ on NVIDIA 1080 GPU. The optimizer was SGD and the initial learning rate is 0.001. We used a batch size of 128 and trained the model for 100 epochs.

\begin{table*}
    \centering
    \caption{Performance of element localisation models.}
    \begin{tabular}{l| c c c |c c c | c c c}
       \hline
       \multirow{2}{*}{\textbf{Method}}   & \multicolumn{3}{c|}{Non-text elements} & \multicolumn{3}{c|}{Text Elements} & \multicolumn{3}{c}{All elements} \\ 
       \cline{2-10}
                                &  \textbf{Precision}  & \textbf{Recall}  & \textbf{F1}  &  \textbf{Precision}  & \textbf{Recall}  & \textbf{F1}   &  \textbf{Precision}  & \textbf{Recall}  & \textbf{F1}  \\
       \hline
       \textbf{UIED}                      & 0.37 & 0.46 & 0.41 & 0.48 & 0.52 & 0.50 & 0.42 & 0.50 & 0.46 \\
       \textbf{FRCNN(nontext)}            & 0.48 & \textbf{0.52} & \textbf{0.50} & / & / & / & / & / & /  \\
       \textbf{FRCNN(nontext) \newline +PaddleOCR}  & 0.47 & 0.48 & 0.47 & 0.56   & 0.48 & 0.51 & 0.54 & 0.50 & 0.52 \\
       \textbf{FRCNN(all)}                & \textbf{0.56} & 0.44 & 0.49 & 0.58 & 0.57 & 0.57 & 0.60 & 0.53 & 0.56 \\
       \textbf{\tool{}}      & \textbf{0.56} & 0.43 & 0.49 & \textbf{0.62} & \textbf{0.62} & \textbf{0.62} & \textbf{0.61} & \textbf{0.55} & \textbf{0.58} \\
       \hline
    \end{tabular}
    \label{tab:localisation_performance}
\end{table*}

\subsection{The Dark Pattern Dataset}
\label{sec:annotation}

To evaluate the overall performance of our proposed system, we need an annotated dataset of dark patterns in UIs that specifies the types and positions of these patterns. However, existing dark pattern datasets are small-scale, incomplete, or only partially labeled. Harry Brignull's Deceptive Design Website provides one example of each included dark pattern type, while Gray et al.'s UXP2 gallery~\cite{gray2018dark} contains 112 artifacts collected from popular online platforms, including UIs from both websites and mobile apps. These datasets are small and do not focus on mobile applications, potentially leading to biased results in evaluation. Di Geronimo et al.~\cite{di2020ui} collected a 10-minute usage video for each of 240 popular Android mobile apps and identified 1,787 dark pattern instances, but only released 15 videos and we were unable to obtain the remaining data from them~\footnote{We emailed the authors but did not get response from them.}. The released videos only contain 105 instances, making them incomplete and small-scale. Additionally, all existing datasets only include labels for the corresponding dark pattern types, but do not specify the location of the patterns within the UIs.

To overcome these limitations, we decided to manually annotate our Rico testset. We chose Rico because it is a relatively comprehensive dataset of UIs collected from a diverse range of apps in different categories and levels of popularity, ensuring the representativeness of the dataset. In total, our test dataset consists of 6,352 UIs from 1,023 apps in 27 app categories.
The annotation process consists of two steps. First, we labelled the types of dark patterns present in each UI. Then, we annotated the positions of these patterns within each UI.

\textit{Annotating the existence and types of dark patterns:} To begin the annotation process, two of the authors studied the integrated taxonomies in Section~\ref{sec:analysis} and examined examples provided by Gray et al. ~\cite{gray2018dark} to gain a shared understanding of dark patterns. To validate this understanding, we used the 15 videos from Di Geronimo et al.~\cite{di2020ui} and independently annotated the existence of dark patterns in each video, discussing and comparing our annotations to those of Di Geronimo et al. Next, each author independently annotated all 6,352 UIs in the Rico testset for the presence of dark patterns and recorded the types of any identified patterns. After completing this step, we discussed our annotations and resolved any discrepancies. This step resulted in the identification of 1,660 instances of 14 dark pattern types from 1,353 UIs and 4,999 benign UIs. As we only identified 14 types of dark patterns in the Rico test dataset, our experiments focused on these types.

\textit{Annotating the locations of dark patterns:}
To ensure consistency in the annotation process, we randomly selected at most 5 UIs from each dark pattern type to serve as standard labeling examples. Based on our analysis in Section~\ref{sec:analysis}, two authors independently annotated the locations of these UIs using the open-source tool LabelImg~\cite{labelImg} and discussed the annotations to form standards for labeling positions. These standards included labeling all relevant UI elements and their element types that may indicate the presence of dark patterns, as well as annotating a container bounding box to encompass the span of the dark pattern. After labeling the sample UIs, we achieved agreement on different annotations and resolved conflicts. The two authors then independently labeled the remaining UIs and resolved any remaining conflicts upon completion.''

\red{We use Cohen's Kappa to evaluate the inter-rater agreements. The agreement for DP presence is 0.97, and for identifying specific types, it is 0.89. These results indicate high agreement between annotators.}

%% file: 7-experiments.tex
\section{Accuracy Evaluation}
\red{In this section, we aimed to evaluate the accuracy of our proposed system to answer two research questions: How accurate is each DL module in our proposed system (RQ1)? How accurate is our proposed system in detecting dark patterns (RQ2)?}

\subsection{RQ1: Accuracy of each DL module}
To evaluate the accuracy of each DL modules in our proposed techniques, we used the corresponding testing datasets as stated in Section~\ref{sec:datasets}.


\subsubsection{Element localisation and classification}
\label{sec:result_localisation}
\textbf{Baseline:}
We considered four baselines: UIED, FRCNN (nontext), FRCNN (nontext) with PaddleOCR results and our ablation model (FRCNN-all). 
\textit{UIED}~\cite{chen2020object} is a state-of-the-art UI element detection technique that combines deep learning and traditional methods to achieve precise localization of UI elements. For text detection, it uses the EAST~\cite{zhou2017east} model and for non-text element detection, it employs a novel old-fashioned method with a top-down coarse-to-fine strategy based on the unique characteristics of UIs and UI elements, along with a CNN classifier to classify the elements. However, our task does not require such high precision in element detection, as end-users will examine the results manually. 
Therefore, we also considered the Faster RCNN model trained to detect only non-text UI elements, referred to as \textit{FRCNN (nontext)} for brevity, and used PaddleOCR to detect text elements and merged them with the non-text element detections, called \textit{FRCNN (nontext) $+$ PaddleOCR}. Lastly, we compared an ablation version of our proposed technique, \textit{FRCNN (all)}, to evaluate the performance of the FasterRCNN-only method.

\textbf{Metric:}
We used precision, recall and F1-score to evaluate the models.
Precision is defined as $TP/(TP$+$FP)$, recall is $TP/(TP$+$FN)$ and F1 score is computed by $(2 \times Precision \times Recall) / (Precision $+$ Recall)$.
A True Positive (TP) refers to a detected element matching the ground truth box and its type is right.
A False Positive (FP) refers to a detected element that does either not match a ground-truth object or fail to predict the right element type.
TP is determined based on the interaction over union (IoU) value of two boxes.
The IoU value is calculated by $interaction\_area/union\_area$ of the two boxes.
If a detected instance has a IoU with the groudtruth box over a preset threshold, and is predicted as a right class, we consider it as a match.
We used the regular IoU threshold 0.5 for our task~\cite{ren2015faster, white2019improving}, as the final results would be examined to a human. 
In addition, we note that OCR engine (both EAST and PaddleOCR) will detect precise bounding box for text elements that the groundtruth bounding box may contain it and have a low IoU value. 
Therefore, if a text element is contained by a groundtruth text box, we also consider it as a match.

\textbf{Results:}
Table \ref{tab:localisation_performance} shows the performance of all models. Our model performs better than most baselines in most metrics, achieving 0.49 in F1 in non-text element detection, 0.62 in F1 in text element detection and 0.58 in F1 for all elements.
Under a loose requirement of precision, deep learning models are able to achieve better performance on non-text element detection than UIED, especially in UIs with noisy background.
For text-element, PaddleOCR performs similar to EAST used in UIED, both reaching around 0.50 in F1.
After merging deep learning results with PaddleOCR, the performance of both FRCNN(nontext) in detecting non-text elements slightly degrades.
Surprisingly, as we can see from the table, FRCNN(all) can achieve better performance than FRCNN(nontext)$+$PaddleOCR in detecting non-text elements, around 0.06 higher in F1 score. 
We investigated the reasons and found that OCR engines can detect texts on advertisements, and the groundtruth annotations normally do not contain the advertisement components and elements, which degrades the OCR engines.
Our final model, \tool{} confirms the usefulness of the OCR engine, achieving 0.61 in precision, 0.55 in recall and 0.58 in F1 in detecting all elements.


\subsubsection{Icon Semantic Understanding}
For icon recognition module, we did not setup a baseline as this kind of method is mature. Therefore, we adopted the accuracy metric to evaluate our performance.
Our model achieves 0.97 accuracy in testing dataset, and 0.95 in detections from our system.
Details of the classification results for each icon type can be seen in supplementary materials. 


\subsubsection{Status Recognition}
Similarly, we also used accuracy metric to evaluate of our status recognition module. Our model achieves 0.99 accuracy in testing dataset and 0.94 accuracy in detections from our element localisation model.
Details of the classification results for each type can be seen in supplementary materials.

\begin{table}[]
    \centering
    \caption{Detailed performance results for each dark pattern type}
    \resizebox{0.5\textwidth}{!}{%
    \begin{tabular}{p{0.5\linewidth}|c c|c c c}
    \hline
     \textbf{DP Type}    & \textbf{\# of GT} & \textbf{\# of Detection} & \textbf{Precision} & \textbf{Recall} & \textbf{F1}       \\
     \hline
     NG - Pop-up AD    & 178  & 156 & 0.82  & 0.72 & \textbf{0.77} \\
     NG - Pop-up to rate  &  1   &  10 & 0.10  & \textbf{1.00} & 0.18  \\
     NG - Pop-up to upgrade &  9 &  38 & 0.21  & \textbf{0.89} & 0.34 \\
     \hline
     SN - Forced Continuity  &  1 	&   1 & 1.00  & 1.00 & \textbf{1.00} \\
     \hline
     II - Preselection - Checkbox selected  &  146 	& 128 & 0.86  & 0.75 & \textbf{0.80} \\
     II - Preselection - No Checkbox  &  86 	&  73 & 0.78 & 0.66 & 0.72 \\
     II - AM - False Hierarchy  & 273 	& 187 & 0.65  & 0.45 & 0.53 \\
     II - AM - Disguised AD  &  46 	&  94 & 0.24  & 0.50 & 0.33 \\
     II - AM - General Types - small close buttons  &  684 	& 649 & 0.99  & 0.94 & \textbf{0.97} \\
     \hline
     FA - Social Pyramid &  16 	&  28 & 0.50  & \textbf{0.88} & 0.64\\
     FA - Privacy Zuckering &  117 & 97 & 0.80 & 0.67 & 0.73 \\
     FA - General Types - Countdown ads &  4 	&   4 & 0.75  & 0.75 & \textbf{0.75} \\
     FA - General Types - watch AD to unlock feature &  2 	&   3 & 0.67  & 1.00 & \textbf{0.80} \\
     FA - General Types - Pay to avoid Ad &   97 	& 100 & \textbf{0.87}  & \textbf{0.90} & \textbf{0.88} \\
     \hline
     \hline
     \textit{Nagging (NG)} & 188  & 204 	& 0.67 & 0.73  & 0.70 \\
     \textit{Sneaking (SN)}  &  1 	&   1 & 1.00  & 1.00 & 1.00 \\
     \textit{Interface Inference (II)} & 1,235 & 1,131 & 0.85 & 0.77 & 0.81 \\
     \textit{Forced Action (FA)} & 236  & 232 	& 0.79 & 0.78 & 0.79 \\
     \hline
     \hline
     \multicolumn{6}{l}{\textbf{\red{Excluding cases with limited examples (n<10)}}} \\
     \hline
     \red{\textbf{Macro Average}}  & \red{1,643}  & \red{1,512} & \red{0.82} & \red{0.76} & \red{0.79} \\
     \red{\textbf{Micro Average}}  & \red{1,643}  & \red{1,512} & \red{0.84} & \red{0.77} & \red{0.80} \\
     \hline
     \hline
     \multicolumn{6}{l}{\textbf{\red{OVERALL}}} \\
     \hline
     \textbf{Macro Average}  & \textbf{1,660}  & \textbf{1,568} & \textbf{0.83} & \textbf{0.82} & \textbf{0.82} \\
     \red{\textbf{Micro Average}}  & \red{\textbf{1,660}}  & \red{\textbf{1,568}} & \red{\textbf{0.82}} & \red{\textbf{0.77}} & \red{\textbf{0.79}} \\
    \hline
    \end{tabular}
    }
    \label{tab:detailed_results}
\end{table}

\begin{table*}[]
    \centering
    \caption{Overall Performance on detecting dark patterns}
    \resizebox{\textwidth}{!}{%
    \begin{tabular}{l|c c c|c c c|c c c|c c c||c c c}
    \hline
     \multirow{2}{*}{Method}     & \multicolumn{3}{c|}{\textbf{Nagging}}       
                                & \multicolumn{3}{c|}{\textbf{Sneaking}}      
                                & \multicolumn{3}{c|}{\textbf{Interface Inference}}  
                                & \multicolumn{3}{c||}{\textbf{Forced Action}}  
                                & \multicolumn{3}{c}{\textbf{OVERALL}} \\
                                \cline{2-16}
                                & Precision & Recall & F1 
                                & Precision & Recall & F1 
                                & Precision & Recall & F1 
                                & Precision & Recall & F1 
                                & Precision & Recall & F1             \\
    \hline
     \textbf{Base Model (Text-only)}     & 0.18 & 0.06 & 0.09 
                                         & \textbf{1.00} & \textbf{1.00} & \textbf{1.00} 
                                         & 0.53 & 0.06 & 0.12
                                         & 0.78 & 0.69 & 0.73
                                         & 0.62  &  0.45  &  0.49  \\
                                         
     \textbf{$+$ Icon Semantic}            & 0.38$\uparrow$ & 0.22$\uparrow$ & 0.28$\uparrow$ 
                                         & \textbf{1.00} & \textbf{1.00} & \textbf{1.00} 
                                         & 0.53 & 0.06 & 0.12
                                         & 0.78 & 0.69 & 0.73
                                         & 0.67$\uparrow$  &  0.49$\uparrow$  &  0.53$\uparrow$ \\
     \textbf{$+$ Template Matching}        & \textbf{0.67}$\uparrow$ & \textbf{0.73}$\uparrow$ & \textbf{0.70}$\uparrow$ 
                                        & \textbf{1.00} & \textbf{1.00} & \textbf{1.00} 
                                        & \textbf{0.89}$\uparrow$ & 0.59$\uparrow$ & 0.71$\uparrow$ 
                                        & 0.78 & 0.69 & 0.73
                                        & \textbf{0.84}$\uparrow$  &  0.75$\uparrow$  &  0.79$\uparrow$ \\
     \textbf{$+$ Status Recognition}      & \textbf{0.67} & \textbf{0.73} & \textbf{0.70} 
                                        & \textbf{1.00} & \textbf{1.00} & \textbf{1.00} 
                                        & \textbf{0.88} & 0.68$\uparrow$ & 0.77$\uparrow$ 
                                        & \textbf{0.79}$\uparrow$ & \textbf{0.78}$\uparrow$ & \textbf{0.79}$\uparrow$ 
                                        & \textbf{0.84}  &  0.80$\uparrow$  &  0.81$\uparrow$ \\
     \textbf{$+$ Color\&Grouping}         & \textbf{0.67} & \textbf{0.73} & \textbf{0.70} 
                                        & \textbf{1.00} & \textbf{1.00} & \textbf{1.00} 
                                        & 0.85$\downarrow$ & \textbf{0.77}$\uparrow$ & \textbf{0.81}$\uparrow$ 
                                        & \textbf{0.79} & \textbf{0.78} & \textbf{0.79} 
                                        & 0.83  &  \textbf{0.82}$\uparrow$  &  \textbf{0.82}$\uparrow$ \\
     \hline
    \end{tabular}
    }
    \label{tab:overall_results}
\end{table*}

\begin{figure*}
    \centering
    \subfloat[Nag to rate]{
        \label{fig:nagtorate} 
        \includegraphics[width=0.15\textwidth]{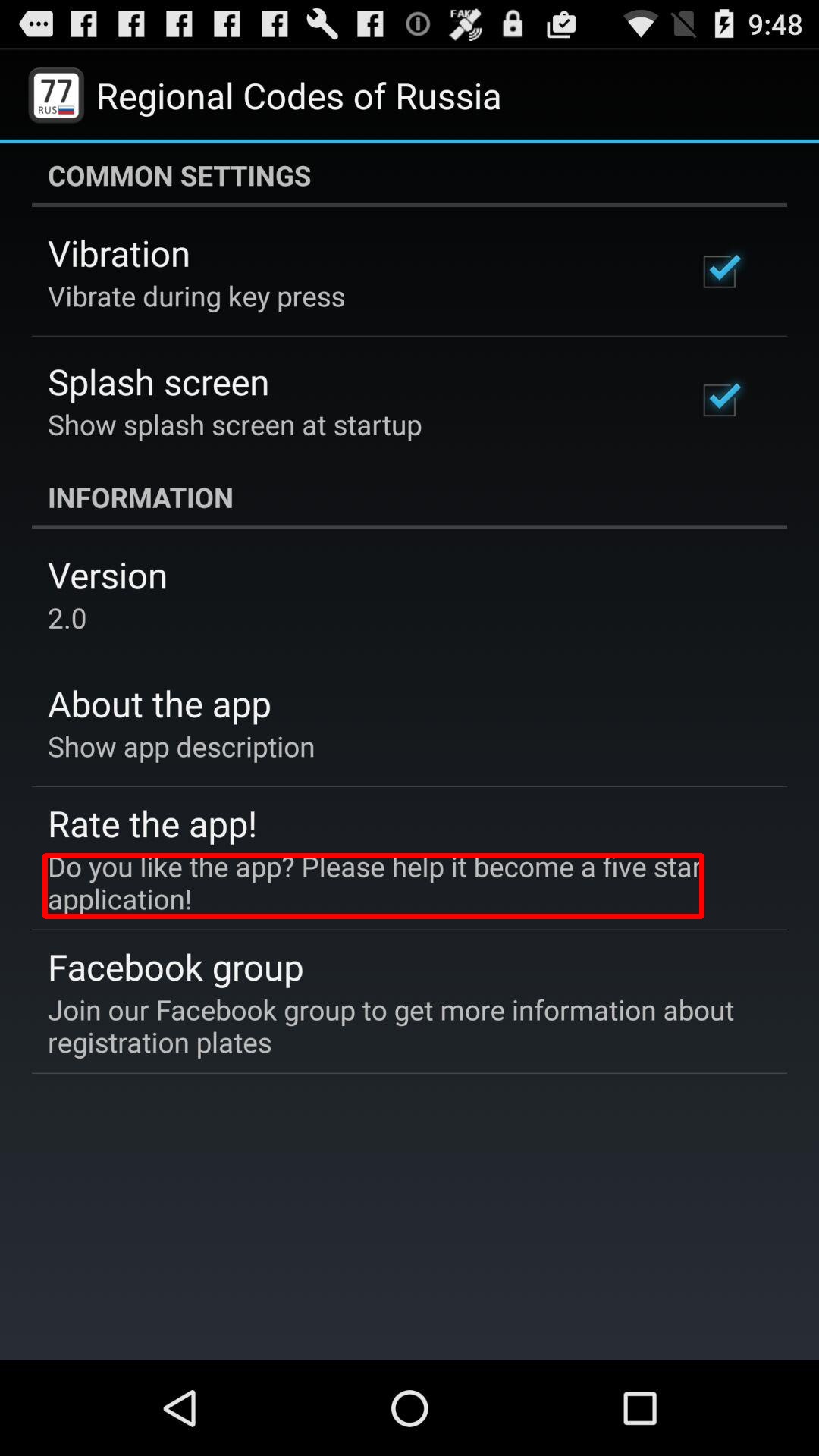}}%
        \Description{A screenshot of a settings UI page featuring several items. One item includes a prompt saying, "Rate the app! If you enjoy our app, please help us reach a five-star rating!"}
    \hspace{0.1cm}
    \subfloat[Nag to upgrade]{
        \label{fig:nagtoupgrade} 
        \includegraphics[width=0.15\textwidth]{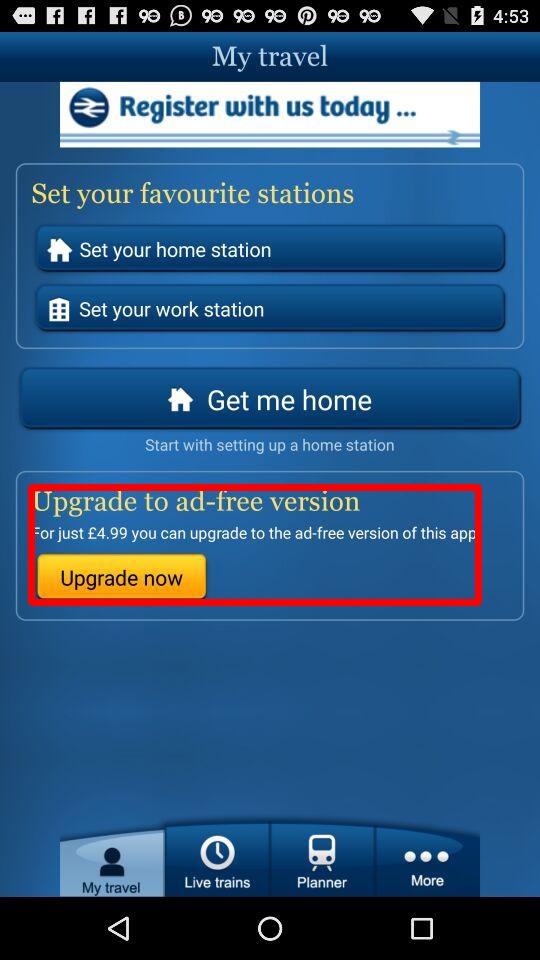}}%
        \Description{This UI features a section at the bottom labeled "Upgrade to ad-free version"}
    \hspace{0.1cm}
    \subfloat[Disguised Ad]{
        \label{fig:disguisedAd} 
        \includegraphics[width=0.15\textwidth]{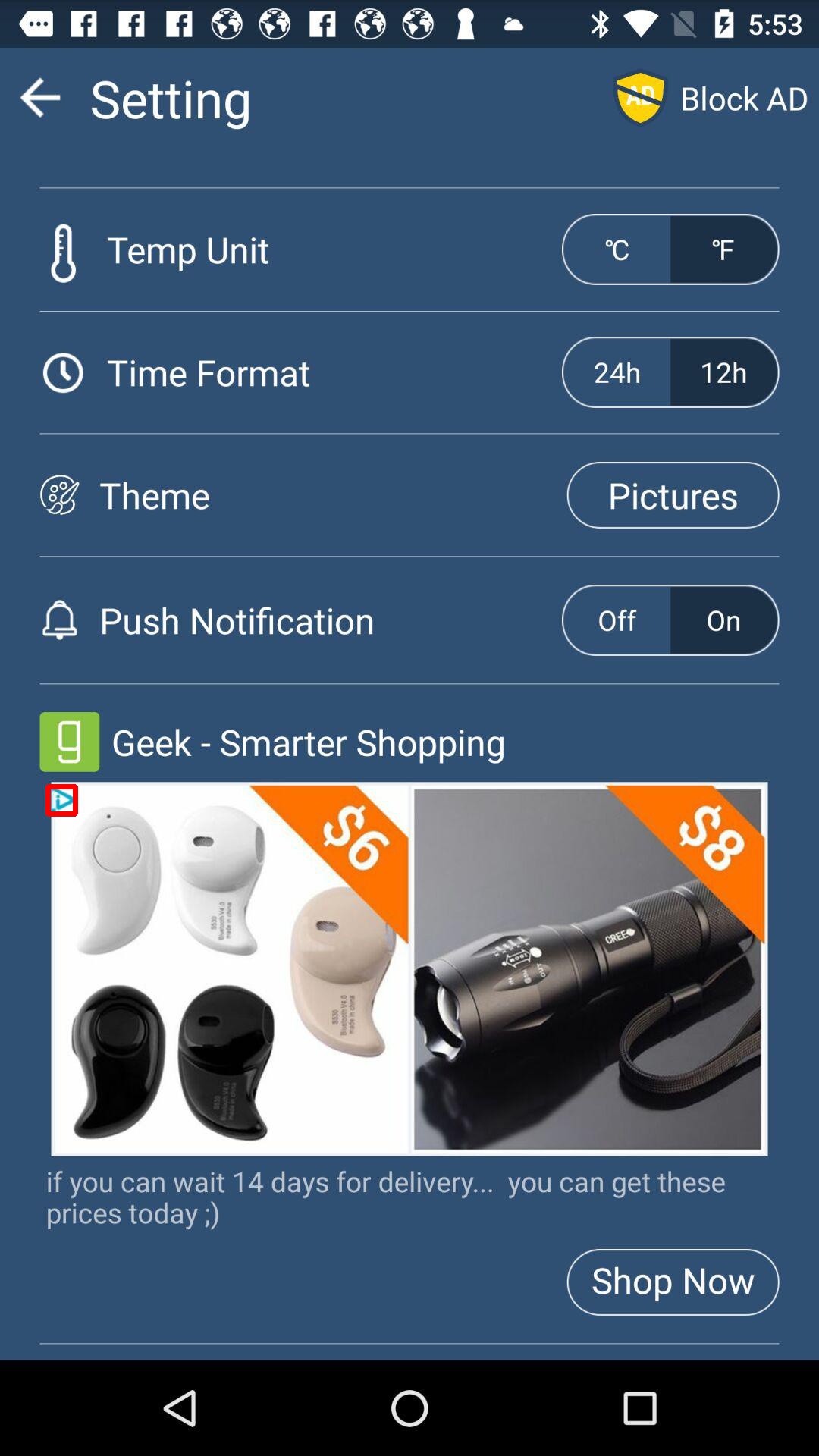}}%
        \Description{The UI contains an embedded ad that noticeably differs from the rest of the content.}
    \hspace{0.1cm}
    \subfloat[Disguised Ad]{
        \label{fig:disguisedAd2} 
        \includegraphics[width=0.15\textwidth]{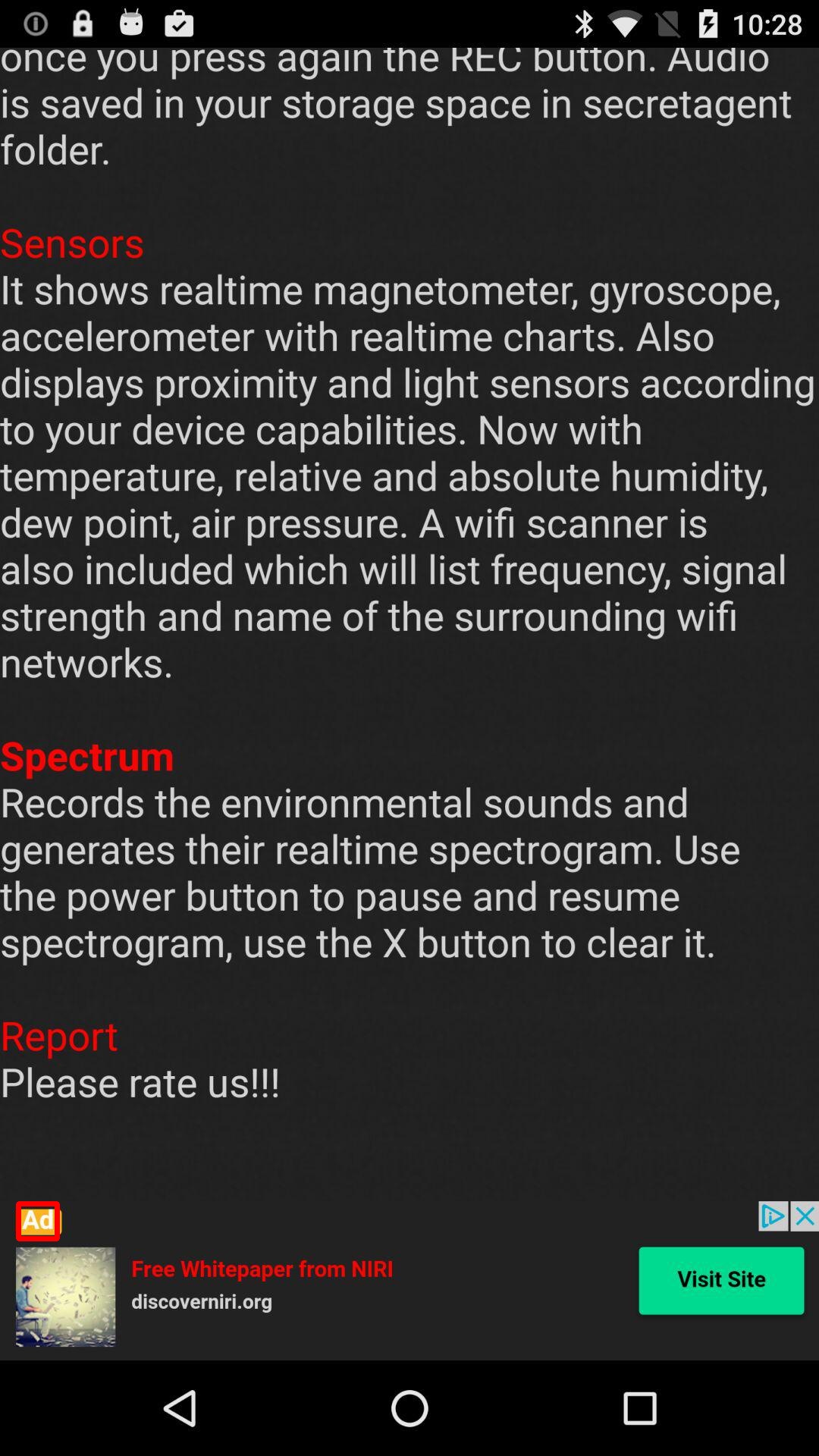}}%
        \Description{The UI includes an advertisement at the bottom, styled similarly to the other content.}
    \caption{Some examples of false positives. \red{(a) A screenshot of a settings UI page featuring several items. One item includes a nagging pattern saying, ``Rate the app! If you enjoy our app, please help us reach a five-star rating!'' (b) A UI features a section at the bottom labeled ``Upgrade to ad-free version.'' (c) A UI contains an embedded ad that noticeably differs from the rest of the content. (d) The UI includes an advertisement at the bottom, styled similarly to the other content. }
    }
    \vspace{-3mm}
    \label{fig:failures}
\end{figure*}

\subsection{RQ2: Accuracy in detecting dark patterns}
\label{sec:overall_accuracy}
We used the dark pattern dataset to evaluate the performance in detecting dark patterns.
Overall, in the dark pattern dataset, \tool{} detects 1,568 instances of dark patterns in 1,304 UIs. 
In terms of the accuracy of classifying whether a UI contains malicious design or not, our system achieves 0.93 accuracy. Of the 4,999 benign UIs, only 192 (3.8\%) are wrongly detected as malicious UIs; Of 1,353 malicious UIs, 82.2\% are correctly detected as UIs with dark patterns.


\subsubsection{Overall Performance}
Table~\ref{tab:detailed_results} shows the detailed results for each dark pattern type, and overall results for each strategy and for all detections. 
\red{We considered two weighted average metric: (1) macro average with equal weights for all categories, and (2) micro average with weights based on category object counts. 
Overall, \tool{} achieves macro/micro average precisions of 0.83/0.82, recall of 0.82/0.77, and F1 score of 0.82/0.79, respectively, which demonstrates that our system can effectively and accurately detect dark patterns in mobile UIs.
To avoid the potential bias from cases with limited examples, we also reported the results after excluding these cases (n<10) and the macro/micro results are 0.82/0.84 for precision, 0.76/0.77 for recall, and 0.79/0.80 for F1. 
}
In terms of strategies, \tool{} also achieves around or over 0.8 for all metrics in Sneaking, Interface Inference and Forced Action strategies.

In terms of the detailed dark pattern types, \tool{} reaches around or over 0.80 in F1 score in 7 types. 
Our system performs well in detecting certain types of dark patterns, such as \texttt{NG-Pop-up AD}, \texttt{II-Preselection - Checkbox selected}, \texttt{II-AM-General Types - small close buttons}, which shows the efficiency of our proposed system.
However, \tool{} performs less well in detecting other types, such as \texttt{NG-Nag to rate} and \texttt{NG-Nag to upgrade}. 
While it remains good recall rates, it detects many false positives, which leads to a low performance in F1 scores. 
We manually checked the reason of false positives in these two types.
As seen in Figure~\ref{fig:nagtorate} and Figure~\ref{fig:nagtoupgrade}, our tool detects in-context rating and upgrade. However, based on the definition of these two dark pattern types, it has to appear as a pop-up window. Some additional techniques such as detecting the existence of pop-up windows can mitigate these kinds of false positives.
In addition, the definition of the \texttt{Disguised AD} dark pattern is ambiguous, leading to poor performance in detecting it.
For example, as seen in Figure~\ref{fig:disguisedAd}, we can see \tool{} detect the potential existence of dark patterns as it locates the ad icon. 
However, as this part of UI is clearly different from other parts as the content is quite different with a big picture, we do not consider it as a disguised ad. Similar situation applies to Figure~\ref{fig:disguisedAd2}.
As people may have different understandings of what is a dark pattern, a more personalized detector that allows users to control which dark patterns they want to be alerted to may be beneficial in addressing this issue."


\subsubsection{Ablation Experiments}
\label{sec:ablation}
We then conducted ablation experiments to understand the usefulness and necessity of each module in our system. To do this, we considered the text-only model as the \texttt{Base Model ($+$Text-only)}, and gradually added \texttt{$+$Icon Semantic} , \texttt{$+$Template Matching}, \texttt{$+$Status Recognition} and \texttt{$+$Color \& Grouping} modules one by one. We calculated the same evaluation metrics for each ablation and noted the trend after the addition of each module.

As we can see in Table~\ref{tab:overall_results}, Base Model performs well in detecting Sneaking and Forced Action dark patterns because these types of dark patterns rely heavily on text content.
In comparison, \texttt{Base Model} does not perform well in detecting Nagging and Interface Inference types, as pure text information is not enough for these types of dark patterns.
After adding the \texttt{Icon Semantic} module, we can see an increase in performance in the Nagging type from 0.09 to 0.28 in F1. 
The \texttt{Template Matching} boosted performance further, as many dark patterns are advertisement related, and identifying the existence of advertisement can be very useful in detecting ad-related dark patterns.  This also means the basic text content and icon semantic modules are insufficient for these types, and our template matching could mitigate errors inherited from the element detection module.
The \texttt{Status Recognition} module specifically deals with checkbox-related dark patterns. While maintaining the precision, its recall goes higher from 0.75 to 0.80 for the overall performance. 
Finally, after adding the \texttt{Color \& Grouping} module, our system gains the ability to recognise dark patterns related to element relationships and color. 
Our final performance achieves 0.83 in precision, 0.82 in recall and 0.82 in F1 score.
The ablation experiments show the usefulness and necessity of each module in our system.



%% file: 8-user_study.tex
\begin{figure*}
    \centering
    \includegraphics[width=1.0\textwidth]{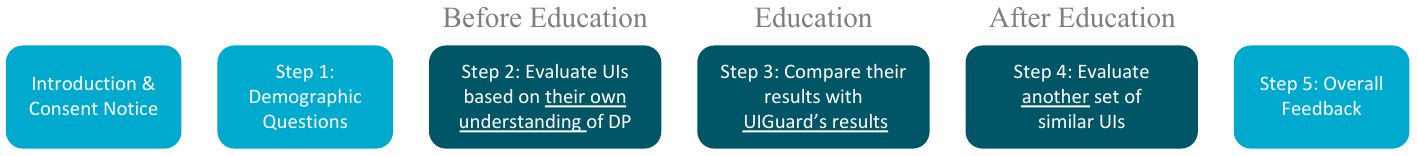}
    \caption{\red{User Study Procedure: Prior to the study, participants was briefed on the study's purpose and procedure, and their consents were obtained. Step 1: Participants completed a demographic questionnaire. Step 2: Participants evaluated UIs based on their dark pattern understanding. Step 3: Participants compared their evaluations with UIGuard's detection results for the UIs from Step 2. Step 4: Participants evaluated a second set of similar UIs. Step 5: Participants provided overall feedback.}}
    \Description{User Study Procedure: Prior to the study, participants was briefed on the study's purpose and procedure, and their consents were obtained. Step 1: Participants completed a demographic questionnaire. Step 2: Participants evaluated UIs based on their dark pattern understanding. Step 3: Participants compared their evaluations with UIGuard's detection results for the UIs from Step 2. Step 4: Participants evaluated a second set of similar UIs. Step 5: Participants provided overall feedback.}
    \label{fig:procedure}
\end{figure*}

\section{Usefulness}
\label{sec:usefulness}

Existing research shows that indicating user the existence of dark patterns can mitigate the potential impacts of them~\cite{di2020ui}. 
In this section, we conducted a user study to measure participants' understanding and perception of dark patterns and the usefulness of \tool{} in detecting and explaining dark patterns. 
\red{We aimed to answer a third research question, i.e., is the proposed system useful from the perspective of ordinary users and how they perceive the existence of dark patterns?}

\subsection{User Study Setup}

\subsubsection{Participants}

To ensure a diverse range of participants, we recruited individuals of various ages, genders, and levels of education through social media platforms such as LinkedIn, Twitter, and alumni networks. All participants were required to be at least 18 years old and have experience using mobile phones. There were no incentives or reimbursements offered for participation in the study.

\subsubsection{Procedure}
The study was conducted in the form of an online survey~\footnote{An example procedure can be seen in \cite{uiguardDataset}}, which consists of five steps \red{(see Figure~\ref{fig:procedure})}. 
Before the survey starts, we gave a short introduction of our user study and asked for the consent from the participants to attend this study.
After obtaining their consent, the first step collected \textit{demographic information} from participants, including their gender, age range, level of education, career, frequency of mobile app usage and their knowledge of dark pattern.

In the second step, participants were asked to evaluate the existence of malicious UI designs in a set of $n$ UIs. They were provided with the definition of a dark pattern (i.e., ``A malicious user interface is defined as a user interface using tricks that make you do things that you didn't mean to''), and were asked to interpret it themselves. If they identified any malicious designs, they were asked to click on the location of the dark pattern on the UI, provide a reason for their identification, and rate the severity and difficulty of the issue on a 5-point Likert scale. Reasons for their ratings were optional to control the length of the study.
The severity is defined as how much impact it will have if they are tricked.
The hardness for recognising the issue is defined as how hard it is to recognize the issue (e.g., how much time it will cost, how carefully the end-user should be).

In the third step, participants were shown the detection results from the tool for each of the k UIs from step 2 (similar to the output in Figure~\ref{fig:approach}), and were asked to evaluate whether they identified the same issues as the tool. If not, they were asked to provide a reason for their discrepancy, rate the severity and difficulty of the issue, and provide their reasons (optional). The participants' own evaluations of the UIs were also shown to them so that they could recall what they recognise in Step 2. 
The reason that we ask the participants to compare their detections with the detection results from our tool, rather than the ground-truth results, is two-fold. First, we want to understand the usefulness of our proposed tool by learning the reasons why people could not identify the issues by themselves. Second, we want to evaluate whether the participants can learn to critique the results from our system and learn something from imperfect AI, as AI systems are not always 100\% correct. This practice can be beneficial during the process of comparing their results with our detections.

Step 4 was a practice session, in which we asked the participants to \textit{evaluate another $m$ UIs with similar issues} as the UIs in Step 2 and asked the same questions as in Step 2. This step aimed to examine whether participants have learned the concept from the short session in step 3, and to understand the educational effect of the tool.

In the final step, participants were asked to provide overall feedback about the usefulness, presentation, advantages, and disadvantages of the tool, and to rate their knowledge of dark patterns again. They were also given the opportunity to provide any other comments. A brief version of the survey is provided in the supplementary materials.
Note that our study passes the ethical assessment from our organisation, and all participants gave their consent on using their answers in our paper.

\subsubsection{Task Selection}

\begin{figure*}
    \centering
    \includegraphics[width=0.8\textwidth]{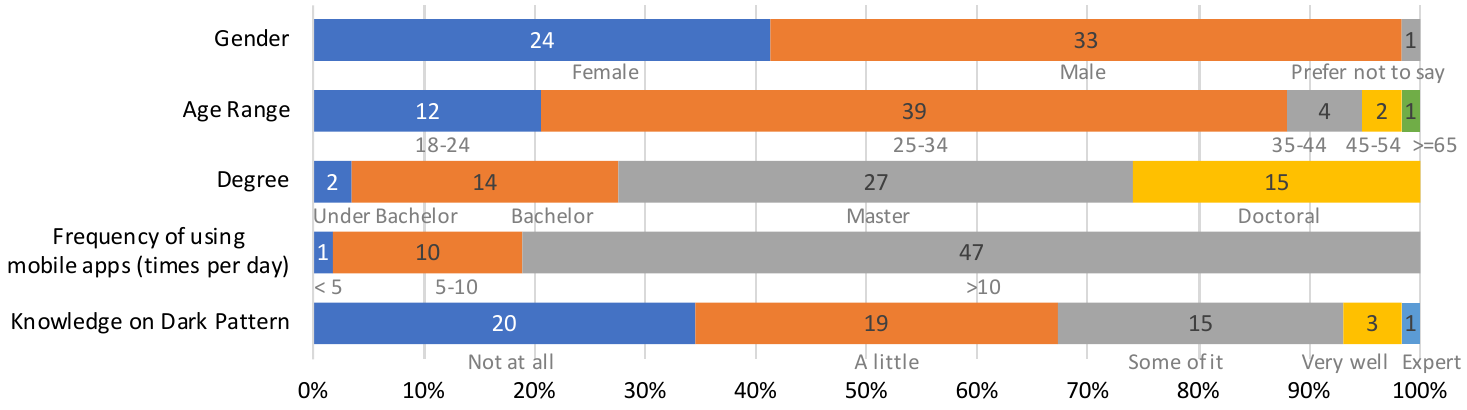}
    \caption{Participants' demographic information}
    \Description{This is a stacked bar chart which contains five rows and shows the 58 participants' demographic information. For top to bottom, the first row is Gender. 24 are female, 33 are male and one prefer not to say. The second row is age range. 12 aged 18-24, 39 aged 25-34, 4 aged 35-44, 2 aged 45-54 and one aged over or equal to 65. The third row is degree. 2 under bachelor, 14 have bachelor degree, 27 have master degree, 15 have doctoral degree. The fourth row is the frequency of using mobile apps per day. 1 uses less than 5 times per day, 10 use 5-10 times per day and 47 use over 10 times per day. The final row is the knowledge on dark pattern. 20 do not have knowledge at all, 19 have a little knowledge, 15 have some of it, 3 have very well knowledge, and 1 considers themselves as experts.
    }
    \label{fig:demographic}
\end{figure*}

For our user study, the participants was required to evaluate $2*n+m$ user interfaces. To control the time consumed by each study (around one hour), we sampled around 20 UIs for Steps 2 and 3, as well as for Step 4. In these sampled UIs, we aimed to include both correct and wrong detections from our tool in order to effectively test its usefulness. Additionally, we also included some benign UIs to evaluate the participants' understanding of ethical and unethical user interfaces.
Given the total of 14 target dark pattern types, we sampled 1-2 instances per type. As such, we adopted a probability-based sampling strategy to ensure a fair representation of different levels of performance for each type of dark pattern from our tool. To align the UIs in Step 4 with Steps 2 and 3, we sampled UIs that were not included in Step 2 and had a similar issue. To minimize bias and ensure representative, we ensured that each UI was sampled from different apps. As a result, we sampled 23 UIs for Step 2 and 3, and 20 UIs\footnote{Excluding UIs that have no dark patterns but are detected by our tool} for Step 4.

To conduct the user study, we used the Qualtrics platform~\footnote{https://www.qualtrics.com/}. Our pilot study indicated that the full survey (in which users need to evaluate in total 66 UIs) would take approximately one hour to complete. As we did not offer any financial incentives to participants, we decided to split the survey into four 15-minute segments. 
For each segment, we ensured that the UIs used in Step 4 were aligned with those used in Step 2. However, we also provided the option for participants to complete the full one-hour version of the survey if they preferred. 
\red{The main distinction between the full survey and the 15-minute versions lies in the number of UIs that participants evaluate in Steps 2, 3, and 4. In the full survey, participants assessed all 23 UIs for Steps 2 and 3, and all 20 UIs for Step 4. Conversely, in the 15-minute versions, participants only evaluated a smaller subset of UIs, 6-7 for Steps 2 and 3, and 5-6 for Step 4. The UIs utilized in the 15-minute versions collectively encompass the UIs employed in the full survey.
All questions and steps except the UIs are the same.}


\subsection{Results}
\subsubsection{Demographic Information}
In total, 58 people participated in our user study.
Of these, 24 finished the full 1-hour version and 34 completed a 15-min questionnaire.
As seen in Figure~\ref{fig:demographic}, the participants were diverse in terms of gender, age, occupation, education level, and mobile phone usage. 
24 of participants were female, 33 were male and 1 preferred not to say.
The ages of the participants ranged from 18 to over 65, with 12 in the 18-24 age range, 39 in the 25-34 age range, 4 in the 35-44 age range, 2 in the 45-54 age range, and 1 over 65. 
Of these participants, 10 were students, 6 were software engineer or developers, 16 were researchers from different domains (e.g., software engineering, Human Computer Interaction (HCI), UI/UX, social, finance and AI), 19 preferred not to say. The rest were freelance, construction worker, data analyst, export operator, or accountant. 
The education levels of the participants varied, with 2 having a degree below a bachelor's degree, 14 having a bachelor's degree, 27 having a master's degree, and 15 having a doctoral degree.
In terms of mobile phone usage, 1 participant used their phone less than 5 times per day, 10 used it 5-10 times per day, and 47 used it more than 10 times per day. 
In terms of knowledge of dark patterns, 20 participants were unaware of dark patterns, 19 had a little knowledge, 15 had some knowledge, 3 knew very well, and 1 considered themselves an expert. These results show that our participants were diverse.
\red{A detailed demographic distribution of each survey can be seen in Section~\ref{sec:detailedDemographic}.}

\begin{table*}[]
    \centering
    \caption{Detailed statistics for user study.}
    \resizebox{0.99\textwidth}{!}{
    \begin{tabular}{l|c c c c c | c c c c c }
    \hline
    \multirow{2}{*}{\textbf{DP type}} & \multicolumn{5}{c|}{\textbf{Step2 - before education}} & \multicolumn{5}{c}{\textbf{Step4 - after education}} \\
    \cline{2-11}
    & \textbf{\# of TP} & \textbf{\# of GT} & \textbf{Recall} & \textbf{Severity (TP$/$FN)} & \textbf{Hardness(TP$/$FN)} & \textbf{\# of TP} & \textbf{\# of GT} & \textbf{Recall} & \textbf{Severity} & \textbf{Hardness}\\
    \hline
    NG-Pop-up AD                        & 7 & 66  & 0.106 & 3.1$/$3.0 & \textbf{1.7$/$3.0} & 36 & 66  & 0.545 & 2.8 & 1.4 \\
    NG-Pop-up to rate                   & 6 & 33  & 0.182 & 2.3$/$2.5 & 2.2$/$2.0 & 6 & 33  & 0.182 & 2.7 & 1.7 \\
    NG-Pop-up to upgrade                & 2 & 66  & \textit{0.03} & \textbf{4.5$/$2.3} & 3.0$/$2.8 & 15 & 33  & 0.455 & 3.0 & 2.1 \\
    \hline
    SN-Forced Continuity                & 4 & 33  & 0.121 & \textbf{4.2$/$1.5} & \textbf{3.2$/$1.5} & 18 & 33  & 0.545 & 3.5 & 2.3 \\
    \hline
    II-Preselection-Checkbox selected   & 15 & 99  & 0.152 & 3.6$/$3.2 & 3.3$/$3.2 & 68 & 132  & 0.515 & 3.2 & 2.5 \\
    II-Preselection-No Checkbox         & 3 & 31  & \textit{0.097} & 3.0$/$3.6 & 3.3$/$3.9 & 26 & 31  & \textbf{0.839} & 3.5 & 2.8 \\
    II-AM-False Hierarchy               & 15 & 66  & 0.227 & \textbf{3.5$/$2.0} & 2.7$/$2.9 & 54 & 99  & 0.545 & 2.6 & 2.4 \\
    II-AM-Disguised AD                  & 40 & 64  & \textbf{0.625} & 3.0$/$3.0 & \textbf{2.5$/$3.4} & 51 & 64  & \textbf{0.797} & 2.9 & 2.5 \\
    II-AM-General Types-small close buttons & 35 & 163  & 0.215 & 3.2$/$2.8 & \textbf{2.2$/$3.2} & 41 & 66  & 0.621 & 3.0 & 2.0 \\
    \hline
    FA-Social Pyramid                   & 0 & 33  & \textit{0.0} & -$/$2.5 & -$/$3.5 & 21 & 33  & 0.636 & 2.9 & 2.1 \\
    FA-Privacy Zuckering                & 16 & 97  & 0.165 & 3.5$/$3.2 & 3.2$/$3.3 & 64 & 97  & 0.660 & 3.3 & 2.6 \\
    FA-General Types-Countdown ads      & 9 & 33  & 0.273 & 3.4$/$3.0 & \textbf{1.4$/$2.4} & 22 & 33  & 0.667 & 3.2 & 2.3 \\
    FA-General Types-watch AD to unlock feature & 8 & 33  & 0.242 & 3.1$/$2.8 & \textbf{2.0$/$3.6} & 14 & 33  & 0.424 & 2.6 & 2.6 \\
    FA-General Types-Pay to avoid Ad    & 3 & 64  & 0.047 & \textbf{4.3$/$2.9} & \textbf{2.0$/$3.2} & 17 & 31  & 0.548 & 3.1 & 2.5 \\
    \hline
    \hline
    \textbf{Overall}  &  163 & 881 & \textbf{0.185} & 3.3$/$2.8 & 2.5$/$3.1  & 453 & 784 & \textbf{0.578} & 3.1 & 2.3 \\
    \hline
    \end{tabular}}
    \label{tab:user_study_metrics}
\end{table*}

\subsubsection{User perception of dark pattern}
When evaluating the presence of malicious UI designs in step 2, participants identified a total of 474 instances. Upon manual review, we found that 163 (34.4\%) of these instances matched with our target dark patterns, resulting in a recall rate of 0.185. 
In detail, for each dark pattern type, we showed the number of correct detection and recall in Table~\ref{tab:user_study_metrics}. ~\footnote{Note that we do not report the number of instances that participants spotted for each type because we cannot tell which type they are.}

Among all dark pattern types, participants performed the best in detecting \texttt{II-AM-Disguised AD}, achieving 0.625 recall rate. 
From their provided reasons, we found that most participants can sense that the advertisements which mixed in the main content are malicious design, but only a few of them can clearly articulate the exact reasons.
However, for other types, recall rates for other types were lower than 0.3, with particularly low rates for \texttt{NG-Pop-up to upgrade}, \texttt{II-Preselection-No Checkbox}, \texttt{FA-Social Pyramid} and \texttt{Pay to avoid AD} (all below 10\%). 
The main reason for these low recall rates was lack of knowledge or awareness of these patterns, with some participants stating that these patterns are common and not a big deal. 
Additionally, some participants believed that \texttt{FA-Social Pyramid} was not a malicious design because users have the option to invite a friend or not.
 

We then reported the average severity and hardness ratings for each dark pattern type and the overall results in Table~\ref{tab:detailed_results}. 
For both scores, we reported two average scores in terms of whether the participants successfully identified the malicious design in Step2 (TP) or were showed by our tool in Step 3 (FN).
Generally, for most dark pattern types, participants had similar severity scores.
For some dark pattern types, including \texttt{NG-Pop up to upgrade}, \texttt{SN-Forced Continuity}, \texttt{II-AM-False Hierarchy}, \texttt{FA-General Types-Pay to avoid Ad}, the average severity scores for TP were much lower than the scores for FN.
This result indicates that participants who identified the dark patterns on their own considered them to be more severe than those shown to them by our tool. 
However, both groups agreed that dark patterns can cause frustration, waste time, and lead to financial loss.
There was more variance in the ratings of difficulty.
For \texttt{NG-Pop up AD}, \texttt{II-AM-Disguised AD}, \texttt{II-AM General Types-small close buttons} and \texttt{FA-General Types-watch AD to unlock feature}, \texttt{FA-General Types-Pay to avoid Ad}, participants who identified these instances independently perceived that it was easier to spot than other participants who can not.
The only reverse pattern lies in \texttt{SN-Forced Continuity}, which was perceived as more difficult by those who were shown it by the tool.


\subsubsection{User capabilities to learn from imperfect AI}
\label{sec:imperfect_AI}

In step 3 of our user study, we included 5 instances where our tool provided incorrect detections of dark patterns. We wanted to see if participants would be able to recognize these incorrect detections and learn from the correct ones. Out of the 92 reasons provided by participants for why they did not spot the wrong detections, only 12 (13\%) recognized that the AI was wrong, 4 (4\%) questioned the detection but were unsure, and the remaining 76 (83\%) accepted the incorrect detections and blamed themselves for not spotting them. 
This highlights the challenge of distinguishing between accurate and inaccurate results from AI systems and the importance of assisting users in evaluating the output of AI systems critically. Fortunately, with our knowledge-driven dark pattern checker, which essentially uses a heuristics-based classifier, we can easily trace the reasoning behind its decisions. For example, in Figure~\ref{fig:approach}, we can see that the system identified point 2 as a false hierarchy instance because it detected two related buttons, one with a grey background and the other with a white background, and the button labeled ``Install'' was highlighted while the button labeled ``No Thanks'' was difficult to notice. Providing this information to end-users can help them understand the decision made by the model and determine if it was correct or not.

\subsubsection{Usefulness of our tool}
\label{sec:knowledge}
In order to evaluate the effectiveness of our approach, we compared the results from Step 2 with those from Step 4. In Step 4, participants identified 570 instances of malicious UI design in the UIs presented to them. Of these, 453 (79.5\%) matched our target dark patterns, yielding a recall rate of 57.8\%.
Overall, the recall rates for most dark pattern types were significantly higher in Step 4 compared to those in Step 2, indicating that participants had learned from the short session in Step 3. However, the recall rates in Step 4 were still relatively low, suggesting that a short session may not be sufficient for participants to fully understand and avoid dark patterns.
Powered with knowledge learned from Step 3, the hardness scores all went lower compared to the results in Step 2/3, while the severity scores were similar as in Step 2/3.

To further assess the impact of our approach, we conducted a Wilcoxon signed-rank test~\cite{wilcoxon1992individual} to determine whether participants' knowledge of dark patterns had increased. The null hypothesis was that participants' knowledge would remain the same, while the alternative hypothesis was that participants' knowledge would increase. The test yielded a p-value of $7.1 \times 10^{-9}$, indicating that the null hypothesis could be rejected and that the difference in knowledge was statistically significant. This suggests that our approach was successful in increasing participants' knowledge of dark patterns.

\begin{figure}
    \centering
\includegraphics[width=0.15\textwidth]{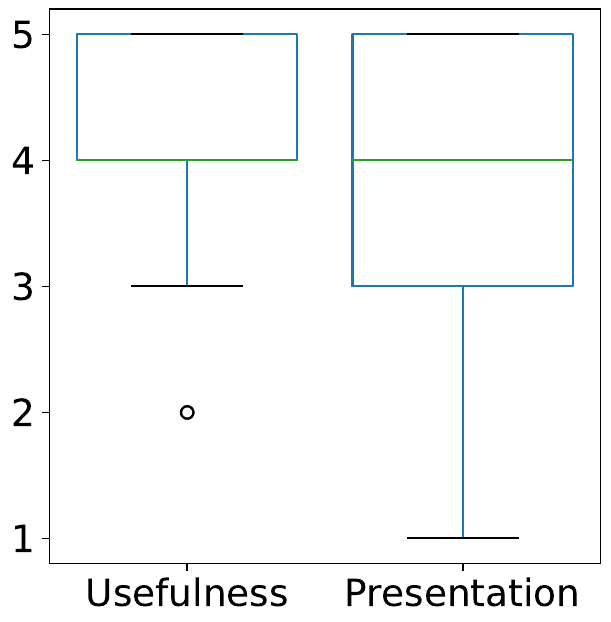}
    \caption{Boxplots for general usefulness and presentation ratings for \tool{}.}
    \Description{These boxplots show the distribution of usefulness and presentation ratings. For both these two aspects, our tool obtains the scores of a median of 4 and the majority of them have a score falling in the range of 3 to 5.}
    \label{fig:general_boxplot}
\end{figure}



Figure~\ref{fig:general_boxplot} shows the boxplots of usefulness and presentation ratings of our tool.
For both these two aspects, \tool{} obtained the scores of a median of 4 and the majority of them had a score falling in the range of 3 to 5, which demonstrates that our system is useful and the current presentation way is acceptable. 
Besides, among all 58 participants, 45 (77.6\%) participants rated the usefulness at 4 or 5, and 42 (72.4\%) of them rated the presentation at 4 or 5.

Generally, participants appreciated the usefulness and accuracy of the tool, and believed this tool could ``\textit{help users detect many dark patterns that may not be easily found by users}'' and ``\textit{avoid them being tricked}'', and thus brought them \textit{better user experience}''. 
They also thought they learned something from this study and their awareness on dark patterns was raised. 
Moreover, some considered this tool can be used as ``\textit{a foundational training tool}'' to raise people awareness in dark patterns, ``\textit{help app stores to examine apps}'' and ``\textit{remind the engineers to enhance their design and think about the ethical concerns}.''
Apart from this, some said they get used to these tricks and could well recognise these tricks by themselves.
One interesting thing is that one participant mentioned that ``even if we know trick, we cannot avoid clicking it sometime.''
In addition, some had different opinions on the concept of certain dark patterns like  ``\textit{Ads are the source of income for some free apps. Watching the ads out of users' choices shouldn't be identified as dark design}.''
In addition, some participants also expressed confusion over distinguishing true detections from false positives.
In terms of presentation, most participants found the tool to be clear and informative, though some mentioned issues with visibility for elderly users.



%% file: 9-discussion.tex
\section{Discussion and Future Work}
\label{sec:discussion}
As the first tool to detect dark pattern in mobile applications, our work has the potentials to benefit a variety of stakeholders, including end-users, app providers (designers and developers), and regulators. 
\red{However, we also emphasise the potential misuse by the app providers, especially if it were actually used by regulators to establish policy. Additional measures should be carefully included to mitigate and avoid this risk.}

\begin{figure}
    \centering
    \includegraphics[width=0.38\textwidth]{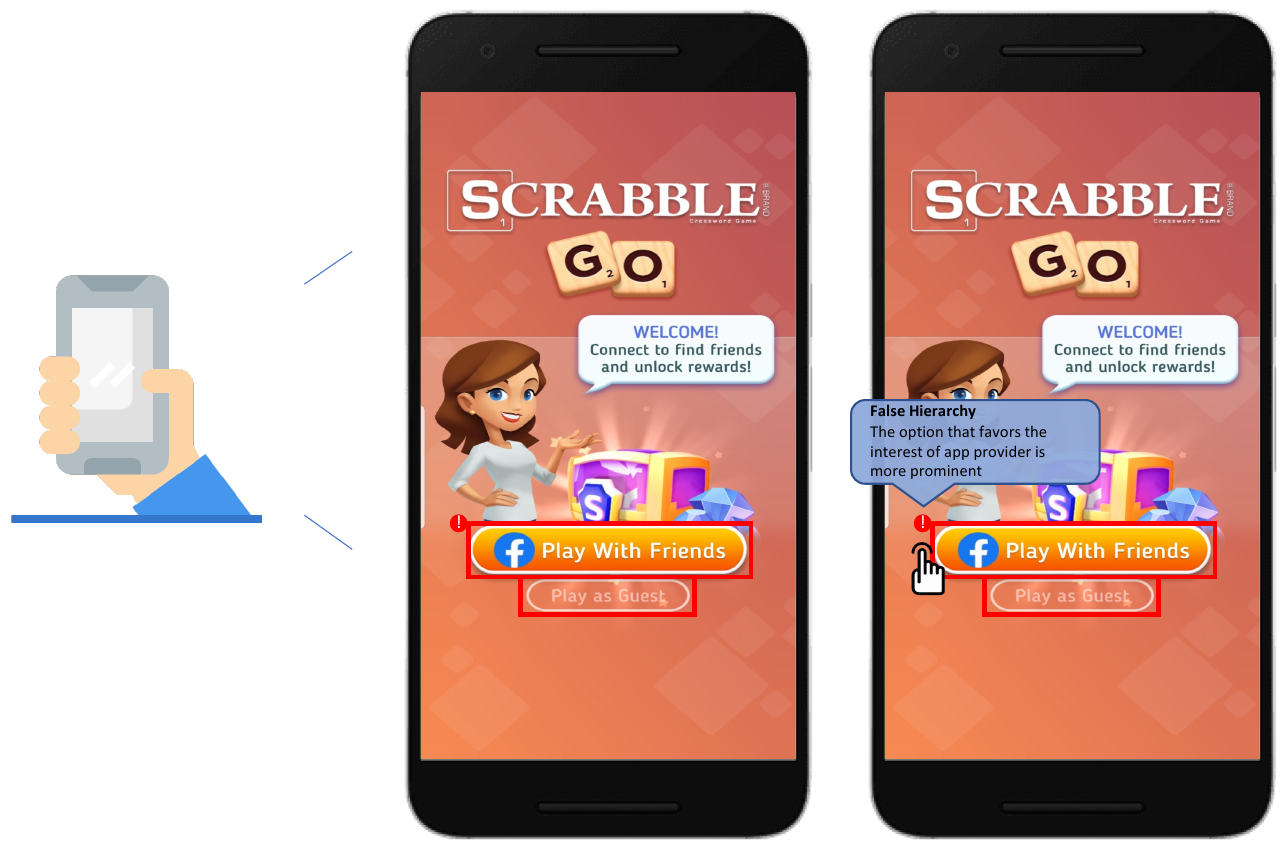}
    \caption{While end-users use the mobile applications, \tool{} can highlight potential dark patterns and avoid them being tricked.}
    \Description{From left to right, there are three figures. The first figure shows a hand holding a mobile phone. The second figure zooms in the screen of the mobile phone, showing a UI page of Scrabble Go. There are two buttons in the middle of the screen, which are highlighted by two separate red boxes, and a question mark lies on the top-left of the red boxes. The third figure shows that when users click the question mark, it shows additional information: "False Hierarchy: The option that favors the interest of app providers is more prominent"}
    \label{fig:app_users}
\end{figure}

\textbf{End-users} are the target users of mobile applications. \tool{} can help end-users to identify and avoid deceptive and manipulative designs in mobile apps, which can improve their user experience and protect them from financial or privacy harm. As seen in Figure~\ref{fig:app_users}, the user is using an app named Scrabble GO, and our \tool{} detects and highlights the presence of dark patterns (in this case, interface inference - aesthetic manipulation - false hierarchy). It also provides detailed information about the specific dark pattern type and its description when the user clicks on the red info icon in the top left corner. This helps users understand the nature of the dark pattern and make informed decisions about how to proceed. However, it is important to consider that not all users may want to receive alerts for every type of dark pattern. Therefore, extra considerations may be necessary when integrating our system into mobile phones.

\begin{figure*}
    \centering
    \includegraphics[width=0.98\textwidth]{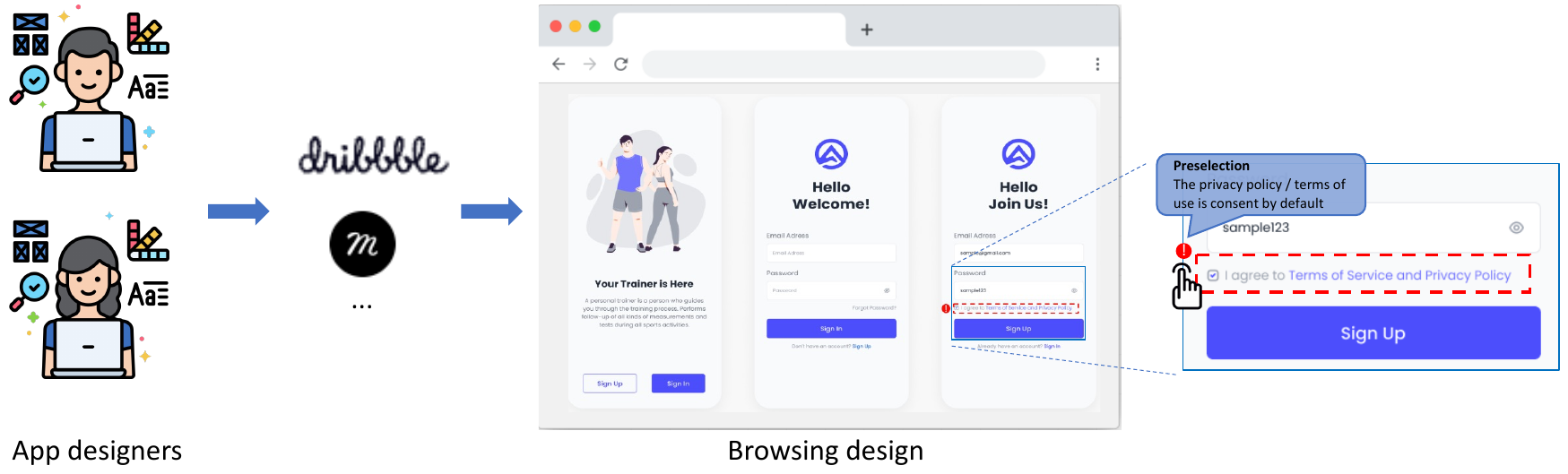}
    \caption{Our system can be embedded into a browser, and automatically detect potential dark patterns in UIs in the design sharing platforms and in case the designers adopt some malicious designs without consciousness.}
    \Description{This figure shows our tool could help highlight the potential dark pattern issues when app designers browse the design in design sharing platforms.}
    \label{fig:app_designers}
\end{figure*}

\textbf{App providers}, including designers and developers, are the ones who design and develop the UIs of apps. 
However, the regulations on dark patterns are often impractical due to their generality, and the various forms of dark patterns, along with the expertise required to recognize them, make it challenging for designers and developers to avoid them. 
In fact, they can even fall victim to dark patterns themselves~\cite{tahaei2021developers}. Therefore, our tool can serve as a guiding tool to improve their design practices and ethical considerations. 
For designers, \tool{} can alert them the potential dark patterns in shared designs in online design sharing platform they are considering using as inspiration. Developers can also use our tool to test for the potential inclusion of dark patterns in their own work. Additionally, providing app providers with clear information about which laws are being violated when dark patterns are detected may increase their desire to avoid using such designs in the future. As seen in Figure~\ref{fig:app_designers}, the designers is browsing the shared design in dribbble platform, and click one design and consider it as trendy and fancy. With the additional assistance of some UI screen extraction techniques~\cite{alahmadi2020uiscreens}, our \tool{} alerts the designers of the dark patterns, and the designer can also learn about dark patterns and gradually recognise the existence of dark patterns by themselves. 
\red{However, there is also a risk that the app providers may misuse such a system to evade dark pattern detection, especially if it were actually used by regulators to establish policy. This potential for misuse underscores the importance of incorporating safeguards. Measures such as continuous algorithmic updates, wider criteria for detection, and stringent regulatory oversight can act as safeguards to prevent this system's misuse, thereby ensuring it continues to serve its original purpose of enhancing user experiences and maintaining app integrity.}

\begin{figure*}
    \centering
    \includegraphics[width=0.8\textwidth]{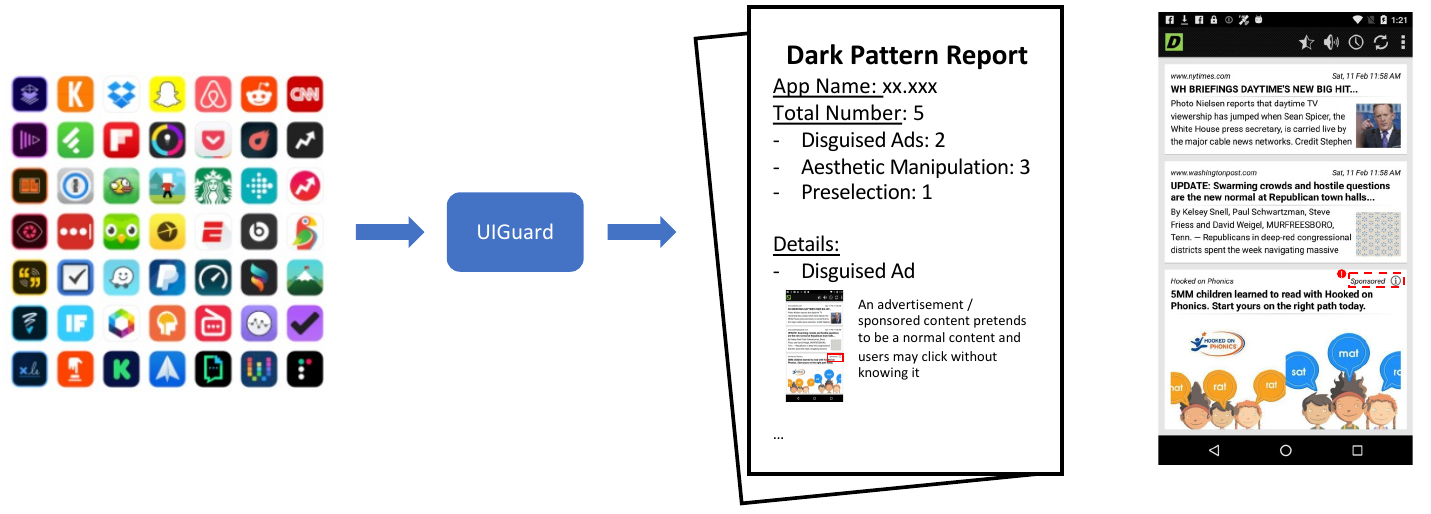}
    \caption{Our system could also be used with an app UI explorer and therefore help regulators to examine the potential existence of dark patterns.}
    \Description{This figure shows that given the app, UIGuard could provide a dark pattern report for regulators to examine the potential violations.}
    \label{fig:app_regulators}
\end{figure*}

\textbf{Regulators.} Many regulations, like EDPB~\footnote{https://edpb.europa.eu/our-work-tools/documents/public-consultations/2022/guidelines-32022-dark-patterns-social-media\_en}, have published some legal documents or guidelines to constrain the usage of dark patterns. However, there are numerous mobile apps in the market. These apps keep updates, introducing new features, and new apps keep coming. It is hard and impossible for the regulators to examine all apps. Paired with some UI exploration tools, our \tool{} can assist regulators in monitoring and enforcing ethical practices in the mobile app market. As seen in Figure~\ref{fig:app_regulators}, based on \tool{} and paired with UI exploration techniques, it can automatically generate a dark pattern report for the regulators even the app markets to easier evaluate apps. \red{Nevertheless, concerns about potential misuse of this technique persist. As such, the strategies highlighted in the preceding paragraph should be actively considered during implementation to guard against such risks.}

Next, we discuss boundary of dark patterns and personalisation, design tradeoff and some limitaitons and future work.


\textbf{Boundary of dark patterns and personalisation.}
The boundary of dark pattern can be a blurry one. While it is important to protect users from deceptive and manipulative design, it is also important to allow for customization and personalization to cater to different user needs. One way to address this issue is to provide adjustable warning levels and allow users to choose which dark patterns to detect and report. This could be done by giving users control over their own settings and allowing them to prioritize certain dark patterns, such as those involving money or privacy. These also are mentioned by our participant in the user study: one said ``people should select what severe results they care most, and the tool should know this and customize its service to different people'', and another noted that ``I think the dark pattern should be prioritized. For example, dark patterns that involve money or privacy should be given more importance by users.'' In the future, it would be interesting to explore ways to balance the need for protection from dark patterns with personalization.


\textbf{Design tradeoff.}
Dark patterns can be harmful to users by deceiving or manipulating them, but detecting and highlighting these patterns can also have negative consequences. In order to embed our system into a mobile phone, we need to use the accessibility service to modify the current user interface, which raises privacy concerns. Additionally, providing extra knowledge or hints through the tool may complicate the user interface, disrupt the app flow, and ultimately annoy users. To mitigate these issues, it may be necessary to consider better methods for integrating the tool and allow users to gradually turn off hints for certain types of dark patterns. False negative reports can also be burdensome, so it is important to consider how to address this issue.

\textbf{\red{Limitations} and Future work.}
As the first work to automatically detect dark patterns in mobile apps, there are several directions we can improve and extend our work:
\red{First, the current work can only detect dark pattern types with signifiers, like disguised ads with ad icons.
Patterns without signifiers may be worse but are not included in our current implementations.
However, based on our findings from the user study (Table~\ref{tab:user_study_metrics} and Section~\ref{sec:usefulness}), participants detected only 62.5\% of disguised ads. With our tool's education, recall rate increased by 27.52\% to 79.7\%. For worse practices without a signifier, we considered them as future work.}
Second, our current work only focused on a subset of dark pattern types, and future work could expand to include more types of dark patterns, particularly dynamic ones.
\red{Third, the taxonomy of dark patterns may evolve when new patterns emerges, our tool can not automatically adapted to the latest changes. Fortunately, our modular method allows easy integration of new rules and property identification modules for emerging DPs. The future work may also focus on the automated evolution of the tool to new concepts. However, we emphasise that monitoring and maintenance are essential for responsible AI practices.}
In addition, the system could be further improved by integrating with UI modification works~\cite{alotaibi2021automated, kollnig2021want} to directly block the malicious designs instead of just highlighting them to users, which could decrease the cognitive burden and protect users from the harms. It is also worth researching on the tradeoff between a direct modification and the hint way.
Moreover, studying the impact of the tool on user behavior, such as whether it causes users to be more cautious when using apps or whether it changes their app usage habits, would be an interesting point.
Lastly, adding more clarification on the legal aspects of dark patterns could also be a valuable direction for future work.



%% file: 10-threats.tex
\section{Threats to Validity}

\subsection{Internal Validity}
\red{
We annotated DPs for evaluation, noticing that the concept's interpretation may vary~\cite{di2020ui, gray2018dark}. Hence, we used existing literature's taxonomy and examples for learning, and individually annotated Di’s videos~\cite{di2020ui} to ensure a consistent and precise grasp of DPs. We also reported the inter-rater agreement results (Cohen's Kappa) in Section~\ref{sec:annotation}. The agreement for DP presence is 0.97, and for identifying specific types, it is 0.89. These results indicate high agreement between annotators, which can mitigate the potential threat.
}

\red{
While our UIGuard performs well in the testing dataset, we acknowledge that some kinds of data may be missed and we were committed to addressing this concern. 
By using the most comprehensive UI dataset, Rico, and carefully annotating the test set in Section~\ref{sec:annotation}, we ensured diverse coverage of UIs. 
Certain DP samples appear a limited number of times, but we actively mitigated potential threats by transparently presenting all statistics in Section~\ref{sec:overall_accuracy} and in Table~\ref{tab:detailed_results}. While the results have slight differences after excluding cases with limited examples, it does not affect the overall findings and conclusions. 
In addition, we released our dataset for future analysis of the potential issues. Adding more dark pattern instances and trendy UIs is in our plans for future work. 
}

\red{
To enhance participation in our study, we segmented the UIs into four subsets, each resulting in a 15-minute survey. This introduced a potential sampling bias as different participants evaluate each survey. We mitigated this by randomly assigning around nine participants per survey. Our demographic analysis in Section~\ref{sec:detailedDemographic} shows no significant differences across surveys. Furthermore, Section~\ref{sec:knowledge}'s evaluation of individual knowledge on dark patterns, unaffected by assignments, affirms UIGuard's effectiveness. 
In addition, additional threat may be introduced due to participant diversity. To counteract this, we sourced participants (n=58) with varied backgrounds from different platforms. 
Hence, we believe these threats are adequately mitigated.
}

\subsection{External Validity}
\red{
Our UIGuard was trained and tested on the Rico and Liu dataset, which was collected in 2017 and 2018. As the UIs keep updating and new design style emerges, we do not yet know the performance of our tool on the latest UIs.
However, Rico dataset remains the most widely-used and high-quality dataset for Android UI designs~\cite{deka2021early}, which can prove the usefulness of our proposed technique to some extent. 
In addition, our approach does not make any specific assumption of UI designs. 
While the style and design rules keep changing over the years, the basic elements remain the same. The UI page always has basic elements, like buttons, images, and icons, which have learned and by our model. 
Our approach is not evaluated on new UIs, but this does not mean our approach and results are fundamentally limited to only those old UIs in Rico.  
Our experiments show that UIGuard achieves around 0.8 F1 scores under different calculations, which is great progress in this domain.
}


%% file: 11-conclusion.tex
\section{Conclusion}
In this work, we analysed existing dark pattern taxonomies for mobile platform, and integrated them into one unified taxonomy.
Our characteristics analysis on the taxonomy categorised the dark pattern into two broad types, i.e., static and dynamic dark patterns, identified six core properties to recognise the existence of dark patterns on screen, and distilled knowledge from practice.
We designed and presented \tool{} that extracts core properties from UI screenshots step by step, and evaluate the existence of dark pattern by pattern matching. Our experiments and user study demonstrate that our \tool{} is promising and useful.
Our work illustrates a novel and systematic approach to mitigate the impacts brought from dark patterns, and many applications could be benefited from this.

%% file: 12-appendix.tex
\section{Detailed Demographic Distribution}
\label{sec:detailedDemographic}
\red{In this section, we show the detailed distributions of all five surveys used in Section~\ref{sec:usefulness}.
We in total have five kinds of survey, one full version (FULL), and four 15-min versions (SHORT1, SHORT2, SHORT3, SHORT4). 
As seen in Table 7-12, no distinct pattern in demographic was found across different segments and the full survey. }

\begin{table}[t]
    \centering
    \begin{tabular}{|l|c|c|c|}
    \hline
        & \textbf{Female} & \textbf{Male} & \textbf{Prefer not to Say} \\
      \hline
      \textbf{FULL}   & 12 & 12 & 0 \\
      \textbf{SHORT1} & 4  & 5  & 0 \\  
      \textbf{SHORT2} & 4  & 4  & 1 \\
      \textbf{SHORT3} & 3  & 4  & 0 \\
      \textbf{SHORT4} & 4  & 5  & 0 \\
      \hline
      \textbf{Total}  & 27 & 30 & 1 \\
      \hline
    \end{tabular}
    \caption{Gender Distribution}
    \label{tab:gender}
\end{table}

\red{\textbf{Gender}: As seen in Table~\ref{tab:gender}, all surveys had an even gender distribution, with near equal representation of male and female participants. }

\begin{table}[t]
    \centering
    \begin{tabular}{|l|c|c|c|c|c|}
     \hline
       & \textbf{18-24} & \textbf{25-34} & \textbf{35-44} & \textbf{45-54} & \textbf{65-74} \\
      \hline
      \textbf{FULL}   & 3 & 19 & 1 & 1 & 0 \\
      \textbf{SHORT1} & 2 & 7 & 0 & 0 & 0 \\  
      \textbf{SHORT2} & 4 & 3 & 1 & 1 & 0 \\
      \textbf{SHORT3} & 1 & 4 & 2 & 0 & 0 \\
      \textbf{SHORT4} & 2 & 6 & 0 & 0 & 1 \\
      \hline
      \textbf{Total}  & 12 & 39 & 4 & 2 & 1 \\
      \hline
    \end{tabular}
    \caption{Age range distribution}
    \label{tab:agerange}
\end{table}

\red{\textbf{Age}: As seen in Table~\ref{tab:agerange}, most prevalent age range was 25-34, followed by 18-24. Fewer participants were in the 35-74 range.}

\red{\textbf{Education Background}: As seen in Table~\ref{tab:education}, majority held a master's degree, with similar numbers of participants having bachelor's or doctoral degrees. Only a small number had qualifications below a bachelor's degree.}

\red{\textbf{Career Distribution}: As seen in Table~\ref{tab:career}, most participants preferred not to disclose their career. The full survey had many researchers and students, while the rest included people from various careers.}

\red{\textbf{Mobile Usage}: As seen in Table~\ref{tab:frequency}, majority of participants reported using mobile apps over 10 times per day, indicating familiarity with mobile applications.}

\red{\textbf{Knowledge on Dark Patterns}:  As seen in Table~\ref{tab:knowledge}, short versions of the survey received varied ratings, ranging from "not at all" to "some" with limited "very well" or "expert" ratings. The full version had more "not at all" and "a little" ratings.}

\begin{table}[t]
    \centering
    \resizebox{0.5\textwidth}{!}{%
    \begin{tabular}{|l|c|c|c|c|}
    \hline
        & \textbf{Under Bachelor} & \textbf{Bachelor} & \textbf{Master} & \textbf{Doctoral} \\
      \hline
      \textbf{FULL}   & 1 & 6 & 12 & 5 \\
      \textbf{SHORT1} & 0 & 3 & 3 & 3 \\  
      \textbf{SHORT2} & 1 & 3 & 2 & 3 \\
      \textbf{SHORT3} & 0 & 2 & 3 & 2 \\
      \textbf{SHORT4} & 0 & 0 & 7 & 2 \\
      \hline
      \textbf{Total}  & 2 & 14 & 27 & 15 \\
      \hline
    \end{tabular}
    }
    \caption{Educational Background Distribution}
    \label{tab:education}
\end{table}

\begin{table}[t]
    \centering
    \resizebox{0.5\textwidth}{!}{%
    \begin{tabular}{|l|c|p{1.25cm}|c|c|p{1cm}|}
    \hline
        & \textbf{Students} & \textbf{Software engineer / developers} & \textbf{researchers} & \textbf{others} & \textbf{prefer} \textbf{not to say} \\
      \hline
      \textbf{FULL}   & 5 & 1 & 8 & 2 & 8 \\
      \textbf{SHORT1} & 1 & 4 & 2 & 0 & 2 \\  
      \textbf{SHORT2} & 3 & 0 & 1 & 1 & 4 \\
      \textbf{SHORT3} & 1 & 0 & 3 & 1 & 2 \\
      \textbf{SHORT4} & 0 & 1 & 3 & 2 & 3 \\
      \hline
      \textbf{Total}  & 10 & 6 & 17 & 6 & 19 \\
      \hline
    \end{tabular}
    }
    \caption{Career distribution.}
    \label{tab:career}
\end{table}

\begin{table}
    \centering
    \begin{tabular}{|l|c|c|c|}
    \hline
        & \textbf{<5} & \textbf{5-10} & \textbf{>10} \\
      \hline
      \textbf{FULL}   & 0 & 2 & 22 \\
      \textbf{SHORT1} & 0 & 3 & 6 \\  
      \textbf{SHORT2} & 0 & 2 & 7 \\
      \textbf{SHORT3} & 1 & 2 & 4 \\
      \textbf{SHORT4} & 0 & 1 & 8 \\
      \hline
      \textbf{Total}  & 1 & 10 & 47 \\
      \hline
    \end{tabular}
    \caption{Frequency of using mobile apps per day (times per day).}
    \label{tab:frequency}
\end{table}

\begin{table}[t]
    \centering
    \resizebox{0.5\textwidth}{!}{%
    \begin{tabular}{|l|c|c|c|c|c|}
    \hline
        & \textbf{Not at all} & \textbf{A little} & \textbf{Some}  & \textbf{Very well} & \textbf{Expert} \\
      \hline
      \textbf{FULL}   & 10 & 10 & 3 & 1 & 0 \\
      \textbf{SHORT1} & 3 & 1 & 4 & 0 & 1 \\  
      \textbf{SHORT2} & 1 & 4 & 3 & 1 & 0 \\
      \textbf{SHORT3} & 2 & 2 & 2 & 1 & 0 \\
      \textbf{SHORT4} & 4 & 2 & 3 & 0 & 0 \\
      \hline
      \textbf{Total}  & 20 & 19 & 15 & 3 & 1 \\
      \hline
    \end{tabular}
    }
    \caption{Knowledge on dark patterns.}
    \label{tab:knowledge}
\end{table}

\cleardoublepage